\documentclass[a4paper,11pt]{article}
\pdfoutput=1
\usepackage{amsmath}
\usepackage{amssymb}
\usepackage{color}
\usepackage{cite}
\usepackage{hyperref}
\usepackage{graphicx}
\usepackage{subfig}

\makeatletter
\@addtoreset{equation}{section}
\renewcommand{\theequation}{\thesection.\@arabic\c@equation}
\makeatother
\makeatletter
\renewcommand\appendix{\par
  \setcounter{section}{0}%
  \setcounter{subsection}{0}%
  \gdef\thesection{Appendix \@Alph\c@section }
  \renewcommand{\theequation}
  {\Alph{section}.\arabic{equation}}
}
\makeatother

\def \be {\begin{equation}}
\def \ee {\end{equation}}
\def \ba {\begin{array}}
\def \ea {\end{array}}
\def \bea{\begin{eqnarray}}
\def \eea{\end{eqnarray}}

\def \b {\beta}

\def \m {\mu}

\def \r {\rho}
\def \o {\omega}

\def \f {\frac}

\def \hs {\hspace}

\def \inf {\infty}

\def \Tr {{\textrm{Tr}}}

\title{\textbf{Large Interval Limit of R\'enyi Entropy At High Temperature}}
\author{
Bin Chen$^{1,2,3}$\footnote{bchen01@pku.edu.cn}\,
and
Jie-qiang Wu$^{1}$\footnote{jieqiangwu@pku.edu.cn}
}
\date{}

\begin{document}

\maketitle

\begin{center}
{\it
$^{1}$Department of Physics, and State Key Laboratory of Nuclear Physics and Technology, Peking University, Beijing 100871, P.R.\! China\\
\vspace{2mm}
$^{2}$Collaborative Innovation Center of Quantum Matter,  \\Beijing 100871, P.~R.~China\\
$^{3}$Center for High Energy Physics, Peking University,  \\Beijing 100871, P.~R.~China
}
\vspace{10mm}
\end{center}

\begin{abstract}


In this paper, we propose a novel expansion to compute the large interval limit of the R\'enyi entropy of 2D CFT at high temperature. Via the replica trick, the  single interval R\'enyi entropy of 2D CFT at finite temperature could be read from the partition function on $n$-sheeted torus connected with each other along a  branch cut. We calculate the partition function by inserting a complete basis across the branch cut. Because of the monodromy condition across the branch cut in the large interval limit, the basis of the states should be the ones in the twist sector. We study the twist sector of a general module of  CFT and find that there is a one-to-one correspondence between the twist sector states and the normal sector states. As an application, we revisit the non-compact free scalar theory and discuss the large interval limit of the R\'enyi entropy of this theory by using our proposal. We find complete agreement in the leading and next-leading orders  with direct expansion of the exact partition function. Moreover, we prove the relation (\ref{th}) between thermal entropy and the entanglement entropy  for a generic CFT with discrete spectrum.

\end{abstract}

\baselineskip 18pt
\thispagestyle{empty}

\newpage

\section{Introduction}

Entanglement entropy is an important notion to encode the active degrees of freedom in  many-body quantum systems\cite{nielsen2010quantum,petz2008quantum}. It is defined to be the von Neumann entropy of reduced density matrix of subsystem $A$
\be
S_A=-\Tr_A \r_A\log \r_A.
\ee
Here the reduced matrix is obtained by smearing over the degrees of freedom of subsystem $B$ complement to $A$
\be
\r_A=\Tr_B \r,
\ee
with $\r$ being the density matrix of the whole system. If the system is in a pure state, one has
\be
S_A=S_B.
\ee
However, if the system is in a thermal state, then its density matrix should be
\be
\r_{th}=e^{-\b H}
\ee
where $H$ is the total Hamiltonian. In the thermal case, there is
\be
S_A \neq S_B
\ee
due to thermal effect. If $A$ is the whole system, then the entanglement entropy is exactly the thermal entropy of the system.

To compute the entanglement entropy, it is convenient to use the so-called R\'enyi entropy, which is defined to be
\be
S_A^{(n)}=-\f{1}{n-1} \log \Tr_A \r_A^n.
\ee
It is easy to see that the entanglement entropy and the R\'enyi entropy are related by
\be
S_A=\lim_{n \to 1} S_A^{(n)},
\ee
if the analytic continuation $n \to 1$ limit is well-defined. This provides a practical way to read the entanglement entropy.

For a quantum field theory which is of infinite degrees of freedom, the entanglement entropy and R\'enyi entropy are quite difficult to compute. It turns out that in higher dimensions the leading contribution to the entanglement entropy is proportional to the area of the boundary of the subsystem. By the replica trick\cite{Callan:1994py}, the R\'enyi entropy in two dimensional quantum field theory can be transformed into calculating the partition function on a higher genus Riemann surface \cite{Calabrese:2009qy}, which equals to a multi-point correlation function of twist operators up to a normalization
\be\label{replica} S_n=-\frac{1}{n-1}\log \frac{Z_n}{Z_1^n}=-\frac{1}{n-1}\log \langle {\cal{T}}^+(u_1){\cal{T}}^-(u_2)...\rangle, \ee
where ...means other twist operators, and $Z_n$ means the partition function for $n$-sheeted surface connecting along the branch cut. Such a computation is usually a formidable task. The exact higher-genus partition function has only been known for the free boson and free fermion. In general, one has to compute the partition function in a well perturbative way. For example, in the two short interval case, one may use the operator product expansion(OPE) of the twist operators to compute the higher genus partition function perturbatively in the order of small cross ratio\cite{Calabrese:2010he}. Recently, by using this expansion, the holographic computation of the two-interval entanglement entropy for the CFT with gravitation dual has been checked beyond the classical level\cite{Zhang,Chen:2013dxa,Perlmutter:2013paa,Chen:2014kja,Beccaria:2014lqa}.

In principle, the partition function for a higher genus Riemann surface can be transformed  into  summing over a series of multi-point correlation functions on the complex plane by cutting and inserting a complete set of state basis of the theory at some cycles.  By modular duality,  the different ways of cutting the Riemann surface should lead to the same result, but the convergence rates in the summations could be different. The simplest example is the partition function on a torus. In this case, the torus could be cut open along the spacial or thermal cycle such that the time or spacial direction is open. The Euclideanized theory could be quantized along the open direction and the partition function is the sum over all of the resulting states. At low temperature, the thermal cycle is longer than the spacial cycle so it is appropriate to quantize the theory along the time direction and read the partition function. On the contrary, at high temperature, the spacial cycle is longer so it is better to quantize along the spacial direction and compute the partition function. In both situations, it is possible to quantize the theory in the other way but the resulting summation series seem to be pathological (slowly convergent), even though the partition function is actually the same for two kinds of quantization. The strategy in computing the torus partition function has been applied to the study of R\'enyi entropy of a single interval on a torus.


For a general 2D CFT, the entanglement entropy of single interval on a circle at finite temperature has only been discussed not long before.
In \cite{Cardy2}, the low temperature case has been investigated. It was pointed out that there was a universal thermal correction for a primary field in a CFT with a mass gap. In \cite{small}, the single interval R\'enyi entropy of a finite temperature CFT with holographic dual has been studied. It has been shown that the holographic computation is in perfect match with the field theory computation. The strategy underlying the field theory computation is that  one can  expand the density matrix (or the partition function) according to the energy of the states. In other words, the density matrix (or the partition function) could be expanded level by level and  only the first few excitations dominate the contributions. In this case, the $n$-genus Riemann surface comes from the $n$ torus being pasted along the spacial interval. Due to the replica symmetry, one may cut each torus along the spacial cycle or thermal cycle, and insert the complete set of state basis to compute the R\'enyi entropy level by level. Similar to the torus partition function case, at low temperature, one may cut along the spacial cycle of each torus and get $n$ cylinder pasting along the interval. After quantizing the theory along the time direction, it is easy to see that the first few excitations give the dominant corrections to the entropy.

\begin{figure}[tbp]
\centering
\subfloat[n sheets Riemann surface]{\includegraphics[width=4cm]{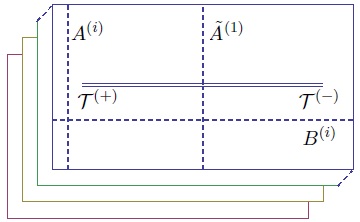}\label{i1}}
\quad
\subfloat[transform the interval]{\includegraphics[width=4cm]{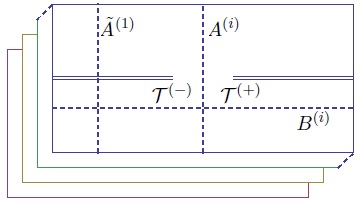}\label{i2}}
\quad
\subfloat[unfold the twist]{\includegraphics[width=4cm]{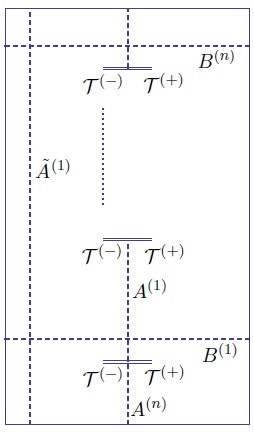}\label{i3}}
\\
\caption{Riemann surface for finite temperature R\'enyi entropy. The horizonal line denotes the spatial direction of unit length, and the vertical line denotes the thermal direction of length $\b$. (a) is the Riemann surface for $n$ cylinders connected by a branch cut which is denoted by a solid line $T_1T_2$. In each sheet, there is a cycle $A^{(i)}$ along the imaginary time direction. However there could also be a cycle marked by $B$, which crosses the branch cut and goes from one sheet to the next one $n$ times until it goes back to the original sheet.   (b) is the same as (a), just by shifting the branch cut to the boundary. In (c) we unfold the twist. The diagram is the example for $n=2$. $T_2T_1$ ($T_2^{'}T_1^{'}$) denotes the complement of the original interval.} 
\end{figure}

The entanglement entropy of a single interval on a circle at high temperature is more subtle. When the interval is not very large, one can use a modular transformation to exchange the thermal direction and spatial direction. One can quantize the theory along the spatial direction rather than the thermal direction such that the thermal density matrix could be taken as $\rho_{th}\propto e^{-RH}=e^{-2\pi (R/ \b) (L_0+\tilde L_0-\frac{c}{12})}$, with $R$ being the spatial length.
Then the partition function (\ref{replica}) could be got by inserting a complete basis along the imaginary time cycles. As the spacial direction becomes open, the torus becomes a cylinder, and the $n$-genus Riemann surface becomes $n$ cylinders connecting along the branch cut.
In Fig. \ref{i1}, the cylinder is unfolded as a rectangle with opposite sides being identified.  In each sheet, there is a cycle $A^{(i)}$ along the imaginary time direction. If the interval is not very large,  one may cut the Riemann surface along the cycle $A^{(i)}$, and insert all the states. For the CFT with pure AdS gravity dual, one can only consider the excitations from the vacuum module in the large central charge limit. The final partition function is an expansion with respect to $e^{-{2\pi RT}}$, as shown in Eqs. (3.14) and (3.29) in \cite{small}. For example, the classical part of the R\'enyi entropy is
\bea\label{class} \lefteqn{S_n\mid_{classical} }\notag \\
&=&\frac{c}{6}\frac{1+n}{n}\log\sinh(2\pi TY)+\mbox{const.}-\frac{c}{9}\frac{(n+1)(n^2-1)}{n^3}\left\{\sinh^4(2\pi Ty)e^{-4\pi TR} \right.\notag\\
&~&+4\sinh^4(2\pi Ty)\cosh^2(2\pi Ty)e^{-6\pi TR}+\left(\frac{-11-2n^2+1309n^4}{11520n^4}\cosh(16\pi Ty) \right.\notag \\
&~&-\frac{-11+28n^2+199n^4}{1440n^4}\cosh(12\pi Ty)-\frac{77-346n^2+197n^4}{2880n^4}\cosh(8\pi Ty) \notag \\
&~&\left.\left.-\frac{-77+436n^2+433n^4}{1440n^4}\cosh(4\pi Ty)+\frac{-77+466n^2+907n^4}{2304n^4}\right)e^{-8\pi TR} \right\} \notag \\
&~&+O(e^{-10\pi TR}), \eea
where $y$ is the length of the interval. When the interval is large enough such that its length is comparable with the size of the circle $y\sim R$, the expansion converge very slowly and is not good anymore. 
This asks us to find another perturbative way to compute the partition function more effectively and reliably.

In this paper, we propose that in the large interval limit, we need another way to cut the Riemann surface and expand the partition function.   In Fig. \ref{i1}, there is also a cycle marked as $B$ which crosses the branch cut and goes from one sheet to the next one $n$ times before it goes back to the original sheet. In Fig. \ref{i2}, we translate the interval and show the cycle $B$ more clearly. We propose to insert a complete basis at the cycle $B$ and compute the R\'enyi entropy perturbatively. As the fields satisfy a certain monodromy condition from one sheet to the next one, we may insert a complete basis in the twist sector and  expand the large interval partition function and the R\'enyi entropy with respect to $e^{-\frac{2\pi RT\Delta_i}{n}}$ order by  order, where $\Delta_i$ is the conformal weight of the primary field of the theory.  After inserting  complete twist sector states,  the calculation at each order  transforms into a four-point function, two operators being the twist operator and anti-twist operator respectively, the other two corresponding to the excited states in the twist sector. In this kind of expansion, the summation series could converge faster.


In the next section, we study the twist sector of orbifold CFT. We show that there is an  one-to-one correspondence between the twist sector states and the normal sector states, with their conformal dimension being related by
\be
h_{twist}=\frac{1}{n}h_{normal}+\frac{c}{24}n(1-\frac{1}{n^2}). \notag\ee
We prove this correspondence first for the vacuum module in a CFT and then a module characterized by an arbitrary primary field.




To check our proposal, we revisit the R\'enyi entropy of a non-compact free scalar at finite temperature in Sec. 3. This case has been discussed before in \cite{Datta}. It was found that the partition function could be written in terms of theta functions. However we notice that the discussion about the relation between $W_1^1$ and $W^2_2$ functions in \cite{Datta} is incorrect. Instead of giving a simple relation between two $W$ functions, we expand them in the small and large interval limits and read the partition functions directly. On the other hand, we recalculate these results by other ways to support our calculation. For the small interval, we use the operator product expansion of the twist operators to compute the partition functions and find a new correction beyond the universal one found in \cite{Cardy2}. We show that this correction comes from continuous spectrum of this gapless model. We find complete agreement with the  result from the $W$ function expansion. In the large interval limit, we need to insert the twist sector states and find that the R\'enyi entropy is in  agreement with the one from direct $W$ functions expansion in the first few orders as well. These good agreements support our treatment on $W$ functions and our proposal on computing the large interval limit of R\'enyi entropy.

Our study of the large interval limit of the R\'enyi entropy on a torus is motivated by the holographic computation of entanglement entropy\cite{Ryu:2006bv,Ryu:2006ef,Lewkowycz:2013nqa}. The holographic entanglement entropy of such a case has been discussed in \cite{Takayanagi,Hubeny:2013gta}. From a holographic point of view, the high temperature 2D CFT is dual to a BTZ black hole. The holographic entanglement entropy of one single interval could be read from the geodesics in the BTZ background with the end being the end points of the interval. When the interval is short, there is nothing special happening, and the geodesic is similar to the one in the global AdS$_3$ and the entropy is just the length of the geodesic. However when the interval is large, there could be two possibilities. One is the usual geodesic length, while the other one could be the sum of the BTZ black hole horizon length and the geodesic length of a very short interval complement to the original one. In other words, when the interval is large, there could be a phase transition. Accordingly the gravitational configuration could be changed from one to the other as the interval increases. Furthermore, it was suggested from the holographic discussion that the thermal entropy and entanglement entropy in 2D field theory is related by
\be\label{th} S_{th}=\lim_{\epsilon\rightarrow 0}(S_{EE}(1-\epsilon)-S_{EE}(\epsilon)). \ee
This relation has been checked in the free fermion case\cite{Takayanagi}. It has also been claimed to be true for the noncompact free boson in\cite{Datta}. 
As an application of our proposal, we prove the relation
for a general CFT with discrete spectrum in Sec. 4.  The key point in our proof is the one-to-one correspondence between the twist sector state and the normal sector state. Moreover, we discuss the relation for the noncompact free scalar. In this case, due to its continuous spectrum, there is a log-logarithmic divergence on the left-hand side of the relation, which is absent in the thermal partition function. This discrepancy could be removed by regularizing the theory and taking the limits appropriately.

We end with some discussions in Sec. 5. In the Appendix, we collect some technical details on  the $W$ functions.

\section{Twist sector}

In this section, we study the twist sector of the orbifold CFT arising from the replica trick in computing the $n$-th Renyi entropy.  As examples, we discuss the vacuum module, and a general CFT module characterized by  a primary operator. The discussions could be extended to a concrete CFT without trouble. There was a similar discussion for the free scalar and fermion in \cite{Dixon0}.

\begin{figure}
  \centering
  \includegraphics[width=7cm]{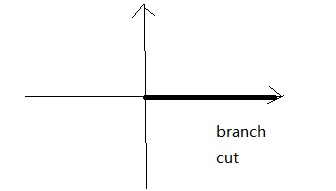}\\
  \caption{One single interval from the origin to the infinity on a full complex plane. }\label{boundary}
\end{figure}

\subsection{Vacuum module}

To simplify the problem, we consider the case that the branch cut is from the origin to infinity. From the replica trick, there are $n$ sheets connected with each other at the branch cut $[0,\inf]$, as in Fig. \ref{boundary}. This results in a $n$-sheeted Riemann surface. Equivalently, we may consider an orbifold CFT on a full complex plane, with $n$ copies of fields and twist boundary conditions around the origin and the infinity. Let us first consider the twist boundary condition around the origin
\be\label{bc} \phi^{(j)}(ze^{2\pi i})=\phi^{(j+1)}(z),  \ee
with $j=1,\cdots, n$ modulo $n$ labeling the sheets. Here $\phi$ can be any field in a CFT.

Let us first consider the vacuum module of a CFT, all of the operators coming from the stress tensor. We have the boundary condition
\be\label{tb} T^{(j)}(ze^{2\pi i})=T^{(j+1)}(z), \ee
and the OPE
\be T^{(i)}(z_1)T^{(j)}(z_2)\sim\delta^{ij}\left(\frac{1}{(z_1-z_2)^4}+\frac{2T^{(j)}(z_2)}{(z_1-z_2)^2}+\frac{\partial T^{(j)}(z_2)}{(z_1-z_2)}\right). \ee
We can redefine the fields as
\be\label{newT} T^{(t,k)}(z)\equiv\sum_{j=1}^{n}T^{(j)}(z)e^{\frac{2\pi i}{n}kj}, \hs{3ex}k=0,1,\cdots,n-1 \ee
  with $T^{(t,0)}$ being the stress tensor for the orbifold CFT, which is the sum over $n$ copies of the original theory's stress tensor. By definition, the monodromy condition is
\be T^{(t,k)}(ze^{2\pi i})=T^{(t,k)}(z)e^{-\frac{2\pi i}{n}k}, \ee
and the OPE for the redefined operators are
\bea T^{(t,k_1)}(z_1)T^{(t,k_2)}(z_2)
&\sim&\delta^{k_1+k_2,rn}\frac{cn}{2}\frac{1}{(z_1-z_2)^4}+\frac{2}{(z_1-z_2)^2}T^{(t,k_1+k_2-rn)}(z_2)\notag \\
&~&+\frac{1}{z_1-z_2}\partial T^{(t,k_1+k_2-rn)}(z_2)+\mbox{(normal ordering)}. \eea
In the above relation, we always choose an integer $r$ such that $0\leqslant k_1+k_2-rn<n$.
We can expand the operators $T^{(t,k)}$ as\footnote{To simplify the notation, we use $L^{(k)}$ to denote the mode expansion of the operators $T^{(t,k)}$ in the twist sector. It does not denote the mode expansion of $T^{(k)}$ in the $k$-th replica. }
\be T^{(t,k)}(z)=\sum_{m\in {\cal Z}}\frac{L^{(k)}_m}{z^{m+2+\frac{k}{n}}}. \ee
For convenience, we may define
\be L_{m+a}^{(k-an)}\equiv L_m^{(k)}, \ee
where $0\leq k<n$ and $a$ is a non-zero integer, so we can write the commutating relations in a simple way
\be [L^{(k)}_{m_1},L^{(-k)}_{m_2}]=\frac{1}{12}nc(m_1+1+\frac{k}{n})(m_1+\frac{k}{n})(m_1-1+\frac{k}{n})\delta_{m_1,-m_2}+
[(m_1+\frac{k}{n})-(m_2-\frac{k}{n})]L^{(0)}_{m_1+m_2}, \ee
and
\be [L^{(k_1)}_{m_1},L^{(k_2)}_{m_2}]=[(m_1+\frac{k_1}{n})-(m_2+\frac{k_2}{n})]L^{(k_1+k_2)}_{m_1+m_2}, \ee
for $k_1+k_2\neq rn$.
With proper combination, this algebra is the same as the Virasora algebra on the homogeneous part but is different on the central extension term.

To study the spectrum, we need to pay attention to the commutators with $L_0^{(0)}$,
\be [L^{(0)}_0,L^{(k)}_m]=-(m+\frac{k}{n})L^{k}_m. \ee
When  the operators $L_m^{(k)}$ act on a state, those with
\be ~ m>0, ~0\leq k<n, ~~
 \mbox{or}~~~  m=0, ~0<k<n, \ee
decrease the state's conformal dimension, so they are annihilation operators; while those with
\be ~m<0, ~0\leq k<n, \ee
increase the state's conformal dimension, so they are creation operators. Therefore we can define the vacuum for the twist sector to be
\bea L^{(k)}_m\mid t\rangle=0 ~~~\mbox{for}&~& m>0, ~0\leq k<n, ~~
 \mbox{or}~~~  m=0, ~0<k<n, \eea
 which is annihilated by all of the annihilation operators and has the lowest conformal dimension. With the creation operators acting on the twist vacuum we can get all of the excited states in the twist sector.

As shown in Fig. \ref{boundary}, besides the origin there is another branch point at the infinity around which we may define the anti-twist sector. Under a conformal transformation
\be\label{trans} \tilde{z}=\frac{1}{z}, \ee
 this branch point is transformed to the origin of the $\tilde{z}$ plane. Then the monodromy condition on the fields
\be T^{(j)}(\tilde{z}e^{2\pi i})=T^{(j-1)}(\tilde{z}) \ee
define the anti-twist sector. Similarly we can introduce  new fields
\be\label{atb} T^{(\tilde{t},k)}(\tilde{z})\equiv \sum_{j=1}^{n}T^{(j)}(\tilde{z})e^{\frac{2\pi i}{n}kj}, \ee
with the monodromy condition
\be T^{(\tilde{t},k)}(\tilde{z}e^{2\pi i})=T^{(\tilde{t},k)}(\tilde{z})e^{\frac{2\pi ik}{n}}, \ee
and the modes expansion
\be T^{(\tilde{t},k)}(\tilde{z})=\sum_{m\in {\cal Z}}\frac{\tilde{L}^{(k)}_m}{\tilde{z}^{m+2-\frac{k}{n}}}. \ee
After defining
\be \tilde{L}^{(k+an)}_{m+a}\equiv \tilde{L}^{(k)}_m, \ee
where $0\leq k<n$ and $a$ is a non-zero integer, we find the commutating relations
\be [\tilde{L}^{(k)}_{m_1},\tilde{L}^{(-k)}_{m_2}]=\frac{1}{12}nc(m_1+1-\frac{k}{n})(m_1-\frac{k}{n})(m-1-\frac{k}{n})\delta_{m_1,-m_2}
+[(m_1-\frac{k}{n})-(m_2+\frac{k}{n})]\tilde{L}^{(0)}_{m_1+m_2}, \ee
and
\be [\tilde{L}^{(k_1)}_{m_1},\tilde{L}^{(k_2)}_{m_2}]=[(m_1-\frac{k_1}{n})-(m_2-\frac{k_2}{n})]\tilde{L}^{(k_1+k_2)}_{m_1+m_2}, \ee
for $k_1+k_2\neq rn$. The commutators with $\tilde{L}_0^{(0)}$ are
\be [\tilde{L}_0^{(0)},\tilde{L}_m^{(k)}]=(-m+\frac{k}{n})\tilde{L}_m^{(k)}. \ee
The operators $\tilde{L}_m^{(k)}$ with
\be ~m>0,~~0 \leq k<n, \ee
are the annihilation operators, while those with
\be ~m<0,~~0 \leq k<n~~~\mbox{or} ~m=0~~0<k<n, \ee
are the creation operators.
The vacuum for the anti-twist sector should satisfy
\be \tilde{L}^{(k)}_m\mid \tilde{t}\rangle=0, ~\mbox{for}~m>0,~0\leq k<n, \ee
and the excited states are generated by $\tilde{L}^{(k)}_m$  with $m<0,~0\leq k<n$ or $m=0,~0<k<n$ acting on the vacuum.

We can define an inner product between a twist sector state at the origin and an anti-twist sector state at the infinity. Actually, the twist and anti-twist boundary condition always appear simultaneously, and the inner product is only well-defined between a twist sector state and its corresponding anti-twist sector state. From Fig. \ref{boundary} and the conformal transformation (\ref{trans}), the fields have the twist boundary condition (\ref{tb}) at the origin and the anti-twist boundary condition (\ref{atb}) at the infinity. We describe the full Riemann surface by two coordinate patches $z$ and $\tilde{z}$. Near the origin of each coordinate patch, we can act the creation operators on the vacuum to build a full twist and anti-twist sector. Note that the creation operators are different at the origins in different coordinate patches. Then the correlation function defines an inner product between the twist sector state and the anti-twist sector state.

Under the conformal transformation (\ref{trans}),
\be T^{(\tilde{t},k)}(\tilde{z})=T^{(t,k)}(z)(\frac{\partial z}{\partial \tilde{z}})^2, \ee
where the Schwarzian term disappears in this conformal transformation.
The modes at the infinity and the ones at the origin are related by
\be \tilde{L}^{(k)}_m\mid_{\tilde{z}}=L^{(k)}_{-m}\mid_z, \ee
where the creation operators at the origin of $\tilde{z}$ coordinate are the annihilation operators at the infinity of the $z$ coordinate.  With these relations, we can define the inner product between a twist sector state and an anti-twist sector state. The definition of the inner product is actually the same as the one based on self-conjugation\cite{Polchinski}.

Up to now, we have built the twist and anti-twist sector and introduced the inner product between the states in two sectors.  With this knowledge, we now show that there is a  one-to-one correspondence between the twist sector state  of the orbifold and the normal sector state in a full complex plane.

To study the relation between the twist sector states and the normal sector states, we need to take a coordinate transformation
\be w=z^{\frac{1}{n}},\label{wz}\ee
which unfolds the $n$-sheeted Riemann surface from Fig. \ref{boundary} to the complex plane $w$. Under this transformation the operator $L^{(k)}_m$ has a correspondent in the $w$ coordinate
\bea\label{trL} L^{(k)}_{m}\mid_z&=&\frac{1}{2\pi i}\oint dzT^{(t,k)}z^{m+\frac{k}{n}+1} \notag\\
&=&\frac{1}{2\pi i}\oint dz\sum_{j=1}^{n}T^{(j)}(z)e^{\frac{2\pi i}{n}kj}z^{m+\frac{k}{n}+1} \notag \\
&=&\frac{1}{2\pi i}\oint_{\mid_n} dzT^{(1)}(z)e^{\frac{2\pi i}{n}k}z^{m+\frac{k}{n}+1} \notag \\
&=&e^{\frac{2\pi i}{n}k}\frac{1}{2\pi i}\oint dwnw^{n-1}[T(w)(\frac{\partial w}{\partial z})^2+\frac{c}{12}\{w,z\}]
(w^n)^{m+\frac{k}{n}+1} \notag \\
&=&e^{\frac{2\pi i}{n}k}[\frac{1}{n}L_{mn+k}+\frac{c}{24}n(1-\frac{1}{n^2})\delta_{mn+k,0}]\mid_w,
\eea
where $\mid_n$ means taking the integral around the origin $n$ times.
Therefore, for each operator in the twist sector, there is a corresponding operator in the untwist coordinate. For $L^{(k)}_m$ $k\neq0~\mbox{or}~m\neq0$, the corresponding operator is $L_{mn+k}$, so the creation or annihilation operators in the $z$ coordinate correspond to the creation or annihilation operators in the $w$ coordinate. For $L^{(0)}_0$ the corresponding operator is $L_0$ plus a constant term which equals the conformal dimension of the twist vacuum. This is in accord with the fact that when we act the corresponding operators  on the vacua in two coordinates
\bea\label{tr0} &~&L^{(0)}_0\mid t\rangle\mid_z=\frac{c}{24}n(1-\frac{1}{n^2})\mid t\rangle\mid_z, \notag \\
&~&(\frac{1}{n}L_0+\frac{c}{24}n(1-\frac{1}{n^2}))\mid 0\rangle\mid_w=\frac{c}{24}n(1-\frac{1}{n^2})\mid 0\rangle\mid_w, \eea
 they read the same result. Here $\mid t\rangle\mid_z$ is the twist sector vacuum, and $\mid 0\rangle\mid_w$ is the normal sector vacuum. Moreover, there are commutation relations
\bea &~&[L_0^{(0)},L_m^{(k)}]=-(m+\frac{k}{n})L_m^{(k)}\mid_z, \notag \\
&~&[L_0,L_{mn+k}]=-(mn+k)L_{mn+k}\mid_w, \eea
for $m\neq0~\mbox{or}~k\neq0$.
Based on the correspondence between the operators in the twist sector and the normal sector, we can build the states in the twist sector and in the normal sector by acting the corresponding creation operators on the vacua, with their conformal dimensions being related by
\be\label{tw} h_{twist}=\frac{1}{n}h_{normal}+\frac{c}{24}n(1-\frac{1}{n^2}). \ee

Furthermore,  considering the fact that the commutating relations in two coordinates keep the same form, we can show that the inner products of corresponding operators on the two sides are  the same, from (\ref{trL}) and (\ref{tr0}). When we take an inner product of two normal states in the vacuum module, we can always move the creation operators to the left and the annihilation operators to the right, and finally find
\be\label{inn} \langle 0\mid f(L_0)\mid 0\rangle\mid_w, \ee
where $f$ is a polynomial of the operator $L_0$. On the other hand, in the $z$ coordinate we can do the similar operation, because the commutation relations are the same  in two coordinates. Therefore  the inner product in the $w$ coordinate can transform into the one in the $z$ coordinate
\be\label{inn1} \langle \tilde{t}\mid f\left(nL_0^{(0)}-\frac{c}{24}n^2(1-\frac{1}{n^2})\right)\mid t\rangle\mid_z. \ee
Based on (\ref{tr0}), the relations (\ref{inn}) and (\ref{inn1}) produce the same answer. In other words, the inner products of the corresponding operators in the twist sector and the normal sector are actually the same. As a result,  the null states in the $w$ coordinate  correspond to the null states in the $z$ coordinate.

 Let us show the null states in the twist sector for vacuum module explicitly. In the vacuum module, the state $L_{-1}\mid 0\rangle$ and its descendants are null states. Correspondingly  in the twist sector, the state $L^{(n-1)}_{-1}\mid t\rangle$ and those with  creation operators acting on it are also null states. To see this, let us compute  the inner product
\bea \langle\tilde{t}\mid L^{(1)}_0L^{(n-1)}_{-1}\mid t\rangle
&=&\langle\tilde{t}\mid L^{(1)}_0L^{(-1)}_{0}\mid t\rangle \notag \\
&=&\langle\tilde{t}\mid(\frac{2}{n}L^{(0)}_0+\frac{1}{12}nc(\frac{1}{n}+1)\frac{1}{n}(\frac{1}{n}-1))\mid t\rangle \notag\\
&=&0.
\eea
In the above calculation, we used the fact that the conformal dimension of the twist vacuum is $h=\frac{c}{24}n(1-\frac{1}{n^2})$.
There is only one state with the same conformal dimension in the anti-twist sector, so we have proved that the inner product with $L^{(n-1)}_{-1}\mid t\rangle$ is always zero, and also for the states with the creation operators acting on it.

Besides, we would like to point out that there may  be extra primary states in the twist sector. Considering the operators $L^{(n-i)}_{-1}$ acting on the twist vacuum, the resulting states have conformal dimensions
\be h=h_v+\frac{i}{n}, \ee
where $h_v$ is the conformal dimension of the twist vacuum.
It is easy to see  all  such states can be annihilated  by the  operators $L_m^{(0)}, m>0$ in the Virasoro algebra.
So $L^{(n-i)}_{-1}\mid t\rangle$ are the primary states, whose corresponding operators are primary operators. Such new primary operators in the twist sector have been discussed before in the free boson case\cite{Polchinski}.

\subsection{Other modules}

We can extend the previous discussion to other modules of a CFT. To simplify the discussion, we consider the module characterized by a primary field $\phi(z,\bar{z})$ with conformal dimension $(h,\bar{h})$. The modular invariance constrains the operator content of a CFT. For rational CFT, including minimal models and WZNW models,  $h-\bar{h}$ in a module has to be an integer \cite{Francesco}. Therefore in the following discussion, we focus on the primary field with $(h-\bar{h})$ being an integer. 
Because of the appearance of both holomorphic and anti-holomorphic sectors, the discussion is more complicated than the discussion on the vacuum module. In the twist sector the monodromy condition is
\be \phi^{(j)}(ze^{2\pi i},\bar{z}e^{-2\pi i})=\phi^{(j+1)}(z,\bar{z}), \ee
and the OPE is
\be T^{(j_1)}(z_1)\phi^{(j_2)}(z_2,\bar{z}_2)\sim \delta^{j_1,j_2}(\frac{h}{(z_1-z_2)^2}\phi^{(j_2)}(z_2,\bar{z}_2)+\frac{1}{z_1-z_2}\partial\phi^{(j_2)}(z_2,\bar{z}_2)), \ee
and anti-holomorphic part
\be \bar{T}^{(j_1)}(\bar{z}_1)\phi^{(j_2)}(z_2,\bar{z}_2)\sim \delta^{j_1,j_2}(\frac{\bar{h}}{(\bar{z}_1-\bar{z}_2)^2}\phi^{(j_2)}(z_2,\bar{z}_2)
+\frac{1}{\bar{z}_1-\bar{z}_2}\bar{\partial}\phi^{(j_2)}(z_2,\bar{z}_2)). \ee
We may redefine the fields
\be \phi^{(t,k)}(z,\bar{z})=\sum_{j=1}^ne^{\frac{2\pi i}{n}kj}\phi^{(j)}(z,\bar{z}) \ee
with the monodromy condition
\be \phi^{(t,k)}(ze^{2\pi i},\bar{z}e^{-2\pi i})=\phi^{(t,k)}(z,\bar{z})e^{-\frac{2\pi ik}{n}}. \ee
The OPEs for the redefined fields are
\be T^{(t,k_1)}(z_1)\phi^{(t,k_2)}(z_2,\bar{z}_2)\sim\frac{h}{(z_1-z_2)^2}\phi^{(t,k_1+k_2-rn)}(z_2,\bar{z}_2)+
\frac{1}{z_1-z_2}\partial\phi^{(t,k_1+k_2-rn)}(z_2,\bar{z}_2), \ee
and
\be \bar{T}^{(t,k_1)}(\bar{z}_1)\phi^{(t,k_2)}(z_2,\bar{z}_2)\sim\frac{\bar{h}}{(\bar{z}_1-\bar{z}_2)^2}\phi^{(t,k_1+k_2-rn)}(z_2,\bar{z}_2)
+\frac{1}{\bar{z}_1-\bar{z}_2}\bar{\partial}\phi^{(t,k_1+k_2-rn)}(z_2,\bar{z}_2), \ee
with $0\leq k_1+k_2-rn <n$.

To find the spectrum of the twist sector, let us start from the primary field on the complex plane $\o$. The primary field is composed of the holomorphic and anti-holomorphic parts
\be \phi(w,\bar{w})=\phi_L(w)\bar{\phi}_R(\bar{w}), \ee
with the mode expansion \cite{Ginsparg:1988ui}
\bea \phi_L(w)&=&\sum_{m=-h+s}\frac{\phi_m}{w^{m+h}}, \notag \\
\bar{\phi}_R(\bar{w})&=&\sum_{\bar{m}=-\bar{h}+\bar{s}}\frac{\bar{\phi}_{\bar{m}}}{\bar{w}^{\bar{m}+\bar{h}}}. \eea
Here $s$ and $\bar{s}$ are integers.  The monodromy conditions for $\phi_L(w)$ and $\bar{\phi}_R(\bar{w})$ are trivial. Under the conformal transformations (\ref{wz}),  these fields in the $z$ coordinate could be expanded as
\be \phi_L(z)=\phi_L(w)(\frac{\partial w}{\partial z})^h=\frac{1}{n^h}\sum_{m=-h+s}\frac{\phi_m}{z^{h+\frac{m}{n}}}, \ee
\be \bar{\phi}_R(\bar{z})=\bar{\phi}_R(\bar{w})(\frac{\bar{\partial} \bar{w}}{\bar{\partial} \bar{z}})^{\bar{h}}
=\frac{1}{n^{\bar{h}}}\sum_{\bar{m}=-\bar{h}+\bar{s}}\frac{\bar{\phi}_{\bar{m}}}{\bar{z }^{\bar{h}+\frac{\bar{m}}{n}}}, \ee
\be \phi(z,\bar{z})=\phi_L(z)\bar{\phi}_R(\bar{z})=
\frac{1}{n^{h+\bar{h}}}\sum_{m=-h+s}\frac{\phi_m}{z^{h+\frac{m}{n}}}
\sum_{\bar{m}=-\bar{h}+\bar{s}}\frac{\bar{\phi}_{\bar{m}}}{\bar{z }^{\bar{h}+\frac{\bar{m}}{n}}}
\ee
with the monodromy condition
\be \phi_L(ze^{2\pi in})=\phi_L(z)e^{-2\pi i(n-1)h}, \ee
\be \bar{\phi}_R(\bar{z}e^{-2\pi in})=\bar{\phi}_R(\bar{z})e^{2\pi i(n-1)\bar{h}}, \ee
\be \phi(ze^{2\pi in},\bar{z}e^{-2\pi in})=\phi(z,\bar{z})e^{2\pi i(n-1)(\bar{h}-h)}=\phi(z,\bar{z}). \ee
This shows the single valuedness of the field in the $n$-sheeted coordinate. In fact,
on an $n$-sheeted Riemann surface, the operators should have trivial monodromy when it goes around the branch point $n$ times. Even though both $\phi_L$ and $\bar{\phi}_R$ do not carry such monodromy, their products do, provided that $h-\bar{h}$ is an integer. 

We can transform the field theory on the $n$-sheeted Riemann surface  to a single sheet orbifold CFT, by rewriting the fields as
\be \phi^{(j)}(z,\bar{z})=\phi(ze^{2\pi i(j-1)},\bar{z}e^{-2\pi i(j-1)}). \ee
In this way we can read the expansion for the field in the twist sector
\bea \phi^{(t,k)}&=&\frac{1}{n^{h+\bar{h}-1}}e^{\frac{2\pi i}{n}k}
\sum_{\substack{m=-h+s,\bar{m}=-\bar{h}+\bar{s} \\m-\bar{m}=k+an}}
\frac{\phi_m\bar{\phi}_{\bar{m}}}{z^{h+\frac{m}{n}}\bar{z}^{\bar{h}+\frac{\bar{m}}{n}}} \notag \\
&=&\frac{1}{n^{h+\bar{h}-1}} \sum_{\substack{m=-h+s,\bar{m}=-\bar{h}+\bar{s} \\m-\bar{m}=k+an}}
\frac{\phi^{'}_m\bar{\phi}^{'}_{\bar{m}}}{z^{h+\frac{m}{n}}\bar{z}^{\bar{h}+\frac{\bar{m}}{n}}} ,\eea
where we define
\bea \phi^{'}_m &\equiv& e^{2\pi i\frac{m}{n}}\phi_m, \\
 \bar{\phi}^{'}_{\bar{m}}&\equiv &e^{-2\pi i\frac{\bar{m}}{n}}\phi_{\bar{m}}. \eea
Taking into the OPE we get the commuting relations
\be [L^{(k_1)}_{m_1},\phi^{'}_{m_2}]=[(m_1+\frac{k_1}{n})(h-1)-\frac{m_2}{n}]\phi^{'}_{m_1n+k_1+m_2}, \ee
\be [\bar{L}^{(\bar{k}_1)}_{\bar{m}_1},\bar{\phi}^{'}_{\bar{m}_2}]=
[(\bar{m}_1+\frac{\bar{k}_1}{n})(\bar{h}-1)-\frac{\bar{m}_2}{n}] \phi^{'}_{\bar{m}_1n+\bar{k}_1+\bar{m}_2}. \ee
Under the conformal transformation (\ref{wz}), the operators in the two coordinates are related by
\bea &~&L^{(k)}_m\mid_z\rightarrow
e^{\frac{2\pi i}{n}k}[\frac{1}{n}L_{mn+k}+\frac{c}{24}n(1-\frac{1}{n^2})\delta_{mn+k,0}]\mid_w, \notag \\
&~&\phi^{'}_{m}\mid_z\rightarrow e^{2\pi im}\phi_{m}\mid_w,\notag \\
&~&\bar{\phi}^{'}_{\bar{m}}\mid_z\rightarrow e^{-2\pi i\bar{m}}\phi_{\bar{m}}\mid_w. \eea
As in the vacuum module we may built a one-to-one correspondence between the operators in the twist coordinate $z$ and the one in the untwist coordinate $w$. The first non-zero state is
\be \phi^{'}_{-h}\bar{\phi}^{'}_{-\bar{h}}\mid t\rangle, \ee
with the conformal dimension $(\frac{h}{n}+\frac{c}{24}(1-\frac{1}{n^2}), \frac{\bar{h}}{n}+\frac{c}{24}(1-\frac{1}{n^2}))$, which corresponds to the primary state $\phi_{-h}\phi_{-\bar{h}}\mid 0\rangle$. All the other states can be built from this state by acting on the creation operators $L_{m}^{(k)}$ and $\bar{L}_m^{(k)}$. In the same way as in the vacuum module, we can prove that there is a one-to-one correspondence between the twist sector states and the normal sector states in a module of the CFT. We can also define an inner product between the twist sector state and anti-twist sector state. Similarly the null states in the twist and normal sectors are also in correspondence. For example, in the $w$ coordinate the state
\be \mid\psi\rangle=(L_{-2}-\frac{3}{2(2h+1)}L_{-1}^2)\phi_{-h}\mid0\rangle \ee
is a null state, if
\be h=\frac{1}{16}(5-c\pm\sqrt{(1-c)(25-c)}. \ee
 Under the conformal transformation, in the $z$ coordinate it corresponds to
\be e^{\frac{-2\pi i}{n}(h+2)}\left(nL^{(n-2)}_{-1}-\frac{3}{2(2h+1)}(nL^{(n-1)}_{-1})^2\right)
\phi^{'}_{-h}\bar{\phi}^{'}_{-\bar{h}}\mid t\rangle, \ee
which is also a null state.

Therefore we have shown that for an $n$-sheeted orbifold the twist sector states have a  one-to-one correspondence with single field normal sector states. Their corresponding conformal dimensions are related by (\ref{tw}). This correspondence could be intuitively understood as follows. If we take a coordinate transformation
\be u=\frac{\beta}{2\pi i}\log z, \ee
transforming the $z$ coordinate into a cylinder of spatial length $\b$. There are $n$ copies of the field with the boundary condition
\be \phi_i(u+\beta)=\phi_{i+1}(u). \ee
The energy is \cite{Cardy2}
\bea\label{en} H&=&\frac{2\pi}{\beta}(L_{twist}+\bar{L}_{twist}-\frac{nc}{12})
=\frac{2\pi}{n\beta}(L_{normal}+\bar{L}_{normal}-\frac{c}{12})=\frac{H_{normal}}{n}. \eea
Here in the first equality we have used the fact that the central charge of the orbifold CFT is $nc$   because there are $n$ copies of the fields. And for the next equality we used the relation (\ref{tw}). The relation (\ref{en}) shows that the energy of a twist sector state is $\frac{1}{n}$ of the one of the normal sector state on the cylinder of spatial length $\beta$. The energy of a twist sector state also equals  the energy of the normal sector state on a cylinder of spatial length $n\beta$. Actually we can take a conformal transformation
\be v=\frac{n\beta}{2\pi i}\log z, \ee
unfolding the $n$ copies of fields onto a cylinder of  spatial length $n\beta$. The state on this cylinder is a normal sector state with the energy being $\frac{1}{n}$ of the one of the state on a cylinder of length $\b$, as shown in (\ref{en}). So the twist sector states in the $z$ coordinate are just the normal states in the $v$ coordinate, after we unfold the twist boundary condition. 

The relation (\ref{tw}) suggests that  with respect to the twist sector states, the thermal correction of a primary field of conformal weight $\Delta$  to the R\'enyi entropy should be expanded in terms of $e^{-2\pi RT \Delta/n}$ at high temperature, rather than $e^{-2\pi RT \Delta}$. More generally, the correction is proportional to $e^{-2\pi RT \Delta_t}$, where $\Delta_t$ is the conformal weight of the excitations in the twist sector. We will see this fact explicitly in the noncompact free scalar case.

\section{Non-Compact scalar at high temperature}

In this section, we study the single interval R\'enyi entropy  on a circle  for non-compact complex scalar at high temperature.  The R\'enyi entropy for the free scalar has been discussed in \cite{Datta}. Based on the results obtained in \cite{Datta}, we carefully calculate the leading order contribution with respect to high temperature expansion in both small interval and large interval limits. For the large interval case, we redo the calculation by inserting the twist sector states at the twist boundary, and find the agreement. This gives a consistent check of our recipe\footnote{In this section, we set the spacial length of the worldseet to be unit, i.e. $R=1$, to simplify the notation.}. 

There are many discussions on multi-point correlation functions of twist operators for compact scalar at zero temperature and finite temperature \cite{Dixon0} \cite{Dixon} \cite{torus1} \cite{torus2}. For non-compact scalar, we just need to set the compact radius to infinity. Let us first give a brief review and introduce the strategy to calculate the multi-point twist operators correlation functions.

As we said, the correlation function with only twist operators equals the partition function of the higher genus Riemann surface. On the Riemann surface there are  non-trivial cycles. For a compact scalar, going around the non-trivial cycle, it satisfies the boundary condition
\be \Phi(\tilde{u})=\Phi(u)+mV, \ee
where $\tilde{u}$ means that the argument $u$ moves around a non-trivial cycle and goes back to the original point, $m$ is an integer-valued complex number, and $V$ is the radius of target space. For each boundary condition, there is one classical solution as the saddle point. At each saddle point, the classical action and its quantum correction contribute to the partition function. Summing over the contribution of all saddle points  gives the full partition function. For a free scalar, there is no interaction term, so the quantum correction is 1-loop exact. Since there is only a quadratic term in the action, if we decompose the field as
\be \Phi=\Phi_{saddle~point}+\Phi_{quantum}, \ee
the action for the quantum fluctuation is the same as the original one, insensitive to the boundary condition. In other words, the quantum correction is the same for every saddle point.

The quantum part of the partition function can be computed by using the Ward identity. Since the quantum fluctuation of the scalar is single-valued in the $n$-sheeted Riemann surface, it is easy to transform back to $n$ copies of scalar on a full complex plane with the boundary condition (\ref{bc}). We can define the scalars as before
\be\label{phib} \Phi^{(t,k)}=\sum_{j=1}^{n}e^{\frac{2\pi i}{n}jk}\Phi^{(j)}, \ee
where $0\leq k<n$. Each newly-defined scalar has a simple monodromy condition. The scalar gets a phase factor $e^{2\pi i\frac{k}{n}}$ or $e^{-2\pi i\frac{k}{n}}$ moving around the branch points, and does not change around non-trivial cycles, so these new defined fields do not couple with each other and we can study them separately. In this section we write $\Phi^{(t,k)}$ as $\Phi$  for short. For each scalar, we have the Green functions
\bea\label{Gr} &~&g(u,u';u_i,\bar{u}_i)=\frac{1}{Z_{k,n}}
\langle-\frac{1}{2}\partial_u\Phi\partial_{u'}\bar{\Phi}{\cal{T}}(u_i)...\rangle, \notag \\
&~&h(\bar{u},u';u_i,\bar{u}_i)=\frac{1}{Z_{k,n}}
\langle-\frac{1}{2}\partial_{\bar{u}}\Phi\partial_{u'}\bar{\Phi}{\cal{T}}(u_i)...\rangle, \eea
``..." means other twist operators. By the monodromy condition around each branch point and the single-valuedness around non-trivial cycles
\be \Delta_{\gamma}=\oint_{\gamma}du\partial \Phi+\oint_{\gamma}d\bar{u}\bar{\partial}\Phi=0, \ee
we can determine the Green function (\ref{Gr}). And with these Green functions, we can read the expectation function for the stress tensor on the Riemann surface
\be \langle T(u)\rangle\mid_{RS}=\frac{\langle T(u){\cal{T}}(u_i)...\rangle}{Z_{k,n}}
=\lim_{u'\rightarrow u}(g(u,u')-\frac{1}{(u-u')^2}). \ee
Using the Ward identity
\be T(u){\cal{T}}(u_i)\sim \frac{h}{(u-u_i)^2}+\frac{1}{u-u_i}\partial_{u_i}{\cal{T}}(u_i), \ee
we get
\be \partial_{u_i}\log Z^{qu}_{k,n}=\lim_{u\rightarrow u_i}((u-u_i)\langle T(u)\rangle-\frac{h}{(u-u_i)}), \ee
where $h=\frac{1}{2}\frac{k}{n}(1-\frac{k}{n})$ \cite{Datta}. Solving the differential equation, we find the partition function $Z_{k,n}$ for the scalar $\Phi^{(t,k)}$. Multiplying all the $Z_{k,n}$ for $0\leq k<n$, we get the full quantum contribution of the partition function. And there is no contribution from classical part in non-compact scalar.

The partition function and the R\'enyi entropy for single interval at finite temperature has been studied in \cite{Datta}. For a non-compact scalar, the  correlation function of two twist operators on the torus is \cite{Datta} \cite{Dixon}
\be\label{partition} \langle {\cal{T}}^{+}(u_1){\cal{T}}^{-}(u_2)\rangle=\frac{Z_n}{Z_1^n}
=\prod_{k=0}^{n-1}\frac{2Im\tau}{\mid \det W(k,n)\mid} \mid\frac{\vartheta_1(u_2-u_1\mid\tau)}{\vartheta_1^{'}(0\mid\tau)}\mid^
{-2\frac{k}{n}(1-\frac{k}{n})}, \ee
where $u_1$ and $u_2$ are the branch points and
\be \det W(k,n)= W_1^1(k,n)W_2^2(k,n)-W_2^1(k,n)W_1^2(k,n). \ee
 These $W$ functions are defined in \cite{Datta}. On the torus, the modular parameter has been chosen to be
 \be \tau=i\beta, \ee
such that the $W$ functions are related by
\be (W_1^1)^*=W_1^1=W_1^2,~~~(W_2^1)^*=-W_2^1=W_2^2. \ee
In the Appendix A we list $W^1_1(k,n)$ and $W^2_2(k,n)$ in (\ref{W11}) and (\ref{W22}) respectively.

However, our treatment on $W_1^1$ and $W_2^2$ is different from the one in \cite{Datta}.
Actually we find that the relation (\ref{relationW}) between $W_1^1$ and $W_2^2$ suggested in \cite{Datta} is problematic. Instead, we compute $W_1^1$ and $W_2^2$ explicitly in both the short interval and large interval limits. The detailed computation can be found in the Appendix A.

 For the non-compact scalar, there is no periodic identification in the target space, so that the $n$ scalars $\Phi^{(t,k)}$ are decoupled and  their partition functions $Z_{k,n}$ with the twist boundary condition could be calculated separately. Similarly, the twist operator can also be decomposed into $n$ decoupled ones
\be {\cal{T}}=\prod_{k=0}^{n-1}{\cal{T}}^{(k)}, \ee
where ${\cal{T}}^{(k)}$ only decide the boundary condition for $\Phi^{(t,k)}$. With this definition, we have
\be\label{partition1} \langle {\cal{T}}^{(k)+}(u_1){\cal{T}}^{(k)-}(u_2)\rangle=\frac{Z_{k,n}}{Z_1}
=\frac{2Im\tau}{\mid \det W(k,n)\mid} \mid\frac{\vartheta_1(u_2-u_1\mid\tau)}{\vartheta_1^{'}(0\mid\tau)}\mid^
{-2\frac{k}{n}(1-\frac{k}{n})}. \ee

\subsection{Small interval}

 Let us first consider the short interval case. In this case, we may expand the functions $W_1^1$ and $W_2^2$ and compute the correlation function of the twist operators  directly. Alternatively, as the interval is short, we can take the operator product expansion (OPE) of the twist operators and calculate each term order by order. We indeed find good agreement.

Here we set the two twist operators at $u_1$ and $u_2$ with $u_2-u_1=l$, as in Fig. \ref{i1}, and the solid line between $u_1$ and $u_2$ is the branch cut. The explicit expansion for $W_1^1$ and $W_2^2$ has been given in (\ref{W11s}) and (\ref{W22s}) respectively. Substituting them into the correlation function (\ref{partition}), we obtain
\bea\label{partitions}
\frac{Z_n}{Z_1^n}&=&\langle {\cal{T}}^+(u_1){\cal{T}}^-(u_2)\rangle \notag \\
&=&l^{-\frac{1}{3}n(1-\frac{1}{n^2})}\left(1+\frac{1}{18}n(1-\frac{1}{n^2})\frac{l^2}{\beta^2}[3\pi \b -\pi^2+24\pi^2e^{-\frac{2\pi}{\beta}}]
 +O(l^3)+O(e^{-\frac{4\pi}{\beta}})\right),\notag \\
\eea
and
\bea S_n=\frac{n+1}{3n}\left(\log l+\frac{1}{6}\frac{l^2}{\beta^2}(\pi^2-3\pi\b
-24\pi^2e^{-\frac{2\pi}{\beta}})\right)
+O(l^3)+O(e^{-\frac{4\pi}{\beta}}),
\eea
\bea S_{EE}=\frac{2}{3}\left(\log l+\frac{1}{6}\frac{l^2}{\beta^2}(\pi^2-3\pi\b
-24\pi^2e^{-\frac{2\pi}{\beta}})\right)
+O(l^3)+O(e^{-\frac{4\pi}{\beta}}).
\eea

From the first line of the correlation function, we can see that at the leading order of $e^{-\frac{2\pi}{\beta}}$, it seems to be  different from Cardy and Herzog's universal thermal correction to the single-interval R\'enyi entropy\cite{Cardy2}. At low temperature, the excited states  contribute to the partition function. In \cite{Cardy2}, it has been shown that for a primary field of dimension $\Delta$ in a CFT with mass gap, its leading thermal contribution is of a universal form  $e^{-{2\pi}{\beta}\Delta}$. At high temperature, we can read the universal correction $e^{-\frac{2\pi}{\beta}\Delta}$ by an $S$-duality. Obviously besides such a universal term, there are other terms in (\ref{partitions}).
The discrepancy actually originates from  the continuous spectrum of the non-compact scalar.
Note that the universal thermal correction found in \cite{Cardy2} is based on the assumption that the CFT has a mass gap, which means that the vacuum is unique\footnote{We would like to thank C.P. Herzog for clarifying  the working assumption in \cite{Cardy2}.}. The non-compact scalar obviously does not belong to this class. For a non-compact scalar, at each order of $e^{-\frac{2\pi}{\beta}}$ we have to integrate all of the continuous spectrum, which changes the form of the partition function. To see this point, we compute the partition function in two different ways.

In the first way, we take the OPE of the two twist operator as the interval is small. Because the operator with non-zero momentum has a zero one-point function on the torus, we only keep the zero momentum terms in the OPE
\be\label{OPE} {\cal{T}}^{+}(u_1){\cal{T}}^{-}(u_2)
=c_nl^{-\frac{1}{3}(n-\frac{1}{n})}(1+\sum_i\frac{1}{12}(1-\frac{1}{n^2})l^2T^{(i)}(0)
+\sum_i\frac{1}{12}(1-\frac{1}{n^2})l^2\bar{T}^{(i)}(0)+O(l^3)), \ee
where $T^{(i)}$ and $\bar{T}^{(i)}$ are the energy-momentum tensors in the $i$-th replica.
We can calculate the expectation values of $T^{(i)}$ from the two-point function on a torus \cite{Francesco}
\be \langle\Phi(u,\bar{u})\bar{\Phi}(u',\bar{u'})\rangle=-2\log\mid\frac{\vartheta_1(u-u'\mid\tau)}{\vartheta_1^{'}(0\mid\tau)}
e^{-\pi\frac{(Im(u-u'))^2}{Im\tau}}\mid^2, \ee
where the factor $2$ is because we are considering a complex scalar. Considering
\be T=-\frac{1}{2}:\partial \Phi\partial\bar{\Phi}: ,\ee
we get the expectation of energy momentum tensor
\bea \langle T\rangle&=& \lim_{u\rightarrow u'} -\frac{1}{2}\langle\partial\Phi(u,\bar{u})\partial\bar{\Phi}(u',\bar{u'})\rangle-\frac{1}{(u-u')^2} \notag \\
&=&\pi i\frac{2}{\tau}-\frac{1}{3}\frac{\vartheta_1^{(3)}(0\mid-\frac{1}{\tau})}{\vartheta_1^{'}(0\mid-\frac{1}{\tau})}
-\frac{\pi}{Im\tau} \notag \\
&=&\frac{\pi}{\beta}-\frac{\pi^2}{3\beta^2}(1-24e^{-\frac{2\pi}{\beta}})+O(e^{-\frac{4\pi}{\beta}}).
\eea
Taking this into (\ref{OPE}), we find
\be\label{resmall} \langle{\cal{T}}^{+}(u_1){\cal{T}}^{-}(u_2)\rangle
=c_nl^{-\frac{1}{3}n(1-\frac{1}{n^2})}(1+\frac{l^2}{6}n(1-\frac{1}{n^2})
[\frac{\pi}{\beta}-\frac{\pi^2}{3\beta^2}+8\frac{\pi^2}{\beta^2}e^{-\frac{2\pi}{\beta}}+O(e^{-\frac{4\pi}{\beta}})]+O(l^3)), \ee
which match with (\ref{partitions}) up to $l^2$ and $^{-\frac{2\pi}{\beta}}$. This suggests that our treatment on the $W$ functions is correct.

The $l^2$ term in (\ref{resmall}) is remarkable. The appearance of such term is a general feature  for a single interval in a finite system\cite{Calabrese:2010he}. This kind of term is absent if we follow the relation (\ref{relationW}) and the expansion (\ref{W22s}).

Next let us compute the partition function in a different way. Instead of using the OPE of two twist operators, we insert a complete basis of $n$ copies of the normal sector to compute the multi-point functions directly. The strategy is similar to the one in \cite{Cardy2,small}, with the difference that the spectrum for non-compact scalar is continuous.  For simplicity, let us consider a real non-compact scalar $X$,
whose spectrum includes the primary states $\mid k\rangle$ and their descendants. The conformal dimension for the state $\mid k\rangle$ is $(\frac{k^2}{2},\frac{k^2}{2})$. At high temperature the theory is quantized along the spatial direction, then the torus partition function is
\be Z_1=\Tr e^{-H}=\int dke^{-\frac{2\pi}{\beta}(k^2-\frac{1}{12})}+O(e^{-\frac{2\pi}{\beta}})
=(\frac{\beta}{2})^{\frac{1}{2}}e^{\frac{\pi }{6\beta}}
+O(e^{-\frac{2\pi}{\beta}}), \ee
where the spatial direction is of unit length, and the $O(e^{-\frac{2\pi}{\beta}})$ comes from the descendant states. For other partition functions $Z_n$, the computation is similar to \cite{Cardy2,small}. Here we take $Z_2$ as an example. In this case
\bea\label{Z2} Z_2&=&\int dk_1dk_2e^{-\frac{2\pi}{\beta}(k_1^2+k_2^2-\frac{1}{6})}
\langle k_1k_2\mid {\cal{T}}^+(u_1){\cal{T}}^-(u_2)\mid k_1k_2\rangle+O(e^{-\frac{2\pi}{\beta}}) \notag \\
&=&e^{\frac{\pi}{3\beta}}\langle 0\mid {\cal{T}}^+(u_1){\cal{T}}^-(u_2)\mid 0\rangle
\int dk_1dk_2e^{-\frac{2\pi}{\beta}(k_1^2+k_2^2)}
\frac{\langle k_1k_2\mid{\cal{T}}^+(u_1){\cal{T}}^-(u_2)\mid k_1k_2\rangle}
{\langle 0\mid{\cal{T}}^+(u_1){\cal{T}}^-(u_2)\mid 0\rangle}+O(e^{-\frac{2\pi}{\beta}}), \notag \\
\eea
where $u_1=-\frac{l}{2}, u_2=\frac{l}{2}$. The vacuum correlation function has already been given in finite temperature R\'entyi entropy on infinite space \cite{Calabrese:2009qy}
\be \langle 0\mid {\cal{T}}^+(u_1){\cal{T}}^-(u_2)\mid 0\rangle=(\frac{\beta}{\pi}\sinh \frac{\pi}{\beta}l)^{-\frac{1}{4}}
=l^{-\frac{1}{4}}(1-\frac{1}{24}\frac{\pi^2l^2}{\beta^2}+O(l^4)).
 \ee
For the correlation function on the state $\mid k_1k_2\rangle$,  we can take a conformal transformation
\be\label{uz} z=e^{\frac{2\pi u}{\beta}}, \ee
to  a complex plane with two twist operators being at $z_1=e^{-\frac{\pi l}{\beta}},~z_2=e^{\frac{\pi l}{\beta}}$. The vertex operators for $\mid k\rangle$ at the origin and the infinity are respectively
\bea \mid k\rangle &\rightarrow& e^{ikX(z)}\mid_{z=0} \notag \\
&\rightarrow&e^{-ikX(z')}z'^{k^2}\bar{z}'^{k^2}\mid_{z'\rightarrow\inf}. \eea
With the vertex operators at the origin and the infinity, we may regard the correlator as a multi-point correlation function on a $n$-sheeted Riemann surface connected at the branch cut $[z_1,z_2]$. Under a conformal transformation
\be y=(\frac{z-z_1}{z-z_2})^{\frac{1}{n}}, \ee
the Riemann surface transforms into a full complex plane. Then we find
\bea \lefteqn{\frac{\langle k_1k_2\mid{\cal{T}}^+(u_1){\cal{T}}^-(u_2)\mid k_1k_2\rangle}
{\langle 0\mid{\cal{T}}^+(u_1){\cal{T}}^-(u_2)\mid 0\rangle}} \notag \\
&=&\frac{\langle k_1k_2\mid{\cal{T}}^+(z_1){\cal{T}}^-(z_2)\mid k_1k_2\rangle}
{\langle 0\mid{\cal{T}}^+(z_1){\cal{T}}^-(z_2)\mid 0\rangle} \notag \\
&=&\lim_{z'\rightarrow \inf, z\rightarrow 0}
\langle \prod_{j=1,2}z'^{k_j^2}\bar{z}'^{k_j^2}
e^{-ik_jX^{(j)}(z',\bar{z}')}e^{ik_jX^{(j)}(z,\bar{z})}\rangle_{n-sheet} \notag \\
&=&\lim_{z'\rightarrow \inf, z\rightarrow 0}
\prod_{j=1,2}z'^{k_j^2}\bar{z}'^{k_j^2}
(\frac{\partial y'^{(j)}}{\partial z'})^{\frac{1}{2}k_j^2}
(\frac{\partial \bar{y}'^{(j)}}{\partial \bar{z}'})^{\frac{1}{2}k_j^2}
(\frac{\partial y^{(j)}}{\partial z})^{\frac{1}{2}k_j^2}
(\frac{\partial \bar{y}^{(j)}}{\partial \bar{z}})^{\frac{1}{2}k_j^2} \notag \\
&~&\cdot\langle \prod_{j=1,2}e^{-ik_jX(y'^{(j)},\bar{y}'^{(j)})}e^{ik_jX(y^{(j)},\bar{y}^{(j)})} \rangle \notag \\
&=&(\cosh\frac{\pi l}{2\beta})^{2(k_1-k_2)^2}.
\eea
Here $y^{(j)}$ comes from multi-valuedness  of the transformation (\ref{uz}). And in the last equation we have used the scalar field correlation function \cite{Polchinski}. Taking these result into (\ref{Z2}), we get
\bea   \lefteqn{\int dk_1dk_2e^{-\frac{2\pi}{\beta}(k_1^2+k_2^2)}
\frac{\langle k_1k_2\mid{\cal{T}}^+(u_1){\cal{T}}^-(u_2)\mid k_1k_2\rangle}
{\langle 0\mid{\cal{T}}^+(u_1){\cal{T}}^-(u_2)\mid 0\rangle} }\notag \\
&=&\frac{\beta}{2}\left(1-\frac{2\beta}{\pi}\log(\cosh\frac{\pi l}{2\beta})\right)^{-\frac{1}{2}} \notag \\
&=&\frac{\beta}{2}(1+\frac{\pi l^2}{8\beta}+O(l^4))
\eea
Taking these terms into the correlation function, we get
\bea
\langle{\cal{T}}^+(u_1){\cal{T}}^-(u_2)\rangle=\frac{Z_2}{Z_1^2}
=l^{-\frac{1}{4}}
(1-\frac{1}{24}\frac{\pi^2l^2}{\beta^2}+\frac{1}{8}\frac{\pi l^2}{\beta}+O(l^4)+O(e^{-\frac{2\pi}{\beta}})). \label{conti}\eea
This is the result for the real scalar. If we consider a complex scalar, we need to take the square of the above result. Then we find the agreement with  (\ref{partitions}) valued at $n=2$ up to leading order of $e^{-\frac{2\pi}{\beta}}$. From the computation, we see that the third term proportional to $\frac{\pi}{\b}$ in (\ref{conti}) comes from the continuous spectrum.

\subsection{Large interval}

In this subsection, we calculate the  correlation function (\ref{partition}) and (\ref{partition1}) in the large interval limit. In this limit, with the $W$ functions given in the Appendix, we obtain the correlation function which is expanded with respect to $e^{-\frac{2\pi}{n\beta}}$.  For simplicity, we only do calculation for a single field $\Phi^{(t,k)}$. For the full theory we just need to multiply the contributions from all the  fields together. As in Fig. \ref{i2}, the dashed line between two twist operators is the complement part of the original interval. Its length is $\epsilon$. We may set the right branch point at $u=0$ and the left one at $u=-\epsilon$. For later convenience we also define $x\equiv e^{-\frac{2\pi\epsilon}{\beta}}$.

Inserting (\ref{W11l}) (\ref{W22l}) and (\ref{Sdual}) into (\ref{partition1}), we find for the complex scalar
\be Z_1=e^{\frac{\pi}{3\beta}}(\frac{\beta}{2})+O(e^{-\frac{2\pi}{\beta}}), \ee
and
\bea\label{Z}
Z_{k,n}&=&C\cdot e^{-\frac{2\pi}{\beta}(\frac{k}{n}(1-\frac{k}{n})-\frac{1}{6})}
\beta^{-\frac{2k}{n}(1-\frac{k}{n})}x^{-\frac{k}{n}}(1-x)^{-\frac{2k}{n}(1-\frac{k}{n})}\cdot \frac{1}{F(\frac{k}{n},1-\frac{k}{n},1,1-x)F(1-\frac{k}{n},\frac{k}{n},1,x)}\notag \\
&~&\cdot\left(1+e^{-\frac{2\pi}{\beta}\frac{k}{n}}x^{-\frac{k}{n}}
\frac{F(-\frac{k}{n},\frac{k}{n},1,1-x)}{F(1-\frac{k}{n},\frac{k}{n},1,1-x)}+e^{-\frac{2\pi}{\beta}(1-\frac{k}{n})}x^{\frac{k}{n}}
\frac{F(2-\frac{k}{n},\frac{k}{n},1,1-x)}{F(1-\frac{k}{n},\frac{k}{n},1,1-x)}\right. \notag \\
&~&\left.
+(1-\frac{k}{n})e^{-\frac{2\pi}{\beta}(1-\frac{k}{n})}x^{\frac{k}{n}}
\frac{F(\frac{k}{n},2-\frac{k}{n},2,x)}{F(\frac{k}{n},1-\frac{k}{n},1,x)}
+\frac{k}{n}e^{-\frac{2\pi}{\beta}\frac{k}{n}}x^{1-\frac{k}{n}}
\frac{F(1-\frac{k}{n},1+\frac{k}{n},2,x)}{F(\frac{k}{n},1-\frac{k}{n},1,x)}
+O(e^{-\frac{2\pi}{\beta}})\right). \notag \\
\eea
We have absorbed other coefficients into the constant $C$.

On the other hand, we may compute the partition function $Z_n$ and $Z_{k,n}$ by inserting a complete basis from the twist sector. We find that the leading and next leading terms in the correlation functions via different ways match with each other.  For the twist sector there is no continuous spectrum.  For a complex scalar, the mode expansion for the twist sector is
\bea &~&\frac{\partial \Phi}{\partial z}=\sum_{m=-\inf}^{\inf}\alpha_{m-\frac{k}{n}}z^{-m-1+\frac{k}{n}} \notag \\
&~&\frac{\partial \bar{\Phi}}{\partial z}=\sum_{m=-\inf}^{\inf}\bar{\alpha}_{m+\frac{k}{n}}z^{-m-1-\frac{k}{n}}, \eea
where we write the mode expansion in full complex coordinate
\be\label{full} z=e^{\frac{2\pi}{\beta}u}. \ee
The first excited states are $\alpha_{-\frac{k}{n}}\mid t\rangle$ and $\bar{\alpha}_{-(1-\frac{k}{n})}\mid t\rangle$. Then we have
\bea\label{mod0} Z_{k,n}&=&\sum_i\langle \bar{t},i\mid{\cal{T}}^{-}(-\epsilon){\cal{T}}^{+}(0)\mid t, i\rangle
e^{-\frac{2\pi}{\beta}(\Delta_i-\frac{1}{6})} \notag \\
&=&e^{-\frac{2\pi}{\beta}(\frac{k}{n}(1-\frac{k}{n})-\frac{1}{6})}
(\langle \bar{t}\mid{\cal{T}}^-(-\epsilon){\cal{T}}^+(0)\mid t\rangle
+\frac{\langle \bar{t}\mid\bar{\alpha}_{\frac{k}{n}}{\cal{T}}^-(-\epsilon){\cal{T}}^+(0)\alpha_{-\frac{k}{n}}\mid t\rangle}
{\langle \bar{t}\mid\bar{\alpha}_{\frac{k}{n}}\alpha_{-\frac{k}{n}}\mid t\rangle}
 e^{-\frac{2\pi}{\beta}\frac{k}{n}} \notag \\
&~&+\frac{\langle \bar{t}\mid\alpha_{1-\frac{k}{n}}{\cal{T}}^-(-\epsilon){\cal{T}}^+(0)\bar{\alpha}_{-(1-\frac{k}{n})}\mid t\rangle}
{\langle \bar{t}\mid\alpha_{1-\frac{k}{n}}\bar{\alpha}_{-(1-\frac{k}{n})}\mid t\rangle}
 e^{-\frac{2\pi}{\beta}(1-\frac{k}{n})} \notag \\
 &~&+\mbox{anti-holomophic terms}+O(e^{-\frac{2\pi}{\beta}})).
 \eea
Since we have already transformed to Fig.\ref{i2}, we change the order of two twist operators. By using the conformal transformation (\ref{full}), we can recast the partition function into the one in the complex plane
\bea\label{mod}
Z_{k,n}
&=&e^{-\frac{2\pi}{\beta}(\frac{k}{n}(1-\frac{k}{n})-\frac{1}{6})}
(\frac{2\pi}{\beta})^{\frac{2k}{n}(1-\frac{k}{n})}x^{\frac{k}{n}(1-\frac{k}{n})}
\langle \bar{t}\mid{\cal{T}}^-(x){\cal{T}}^+(1)\mid t\rangle \notag \\
&~&\left\{1+\frac{\langle t\mid\bar{\alpha}_{\frac{k}{n}}{\cal{T}}^-(x){\cal{T}}^+(1)\alpha_{-\frac{k}{n}}\mid \bar{t}\rangle}
{\langle \bar{t}\mid\bar{\alpha}_{\frac{k}{n}}\alpha_{-\frac{k}{n}}\mid t\rangle
\langle \bar{t}\mid{\cal{T}}^-(x){\cal{T}}^+(1)\mid t\rangle }
 e^{-\frac{2\pi}{\beta}\frac{k}{n}} \right.\notag \\
&~&\left.+\frac{\langle \bar{t}\mid\alpha_{1-\frac{k}{n}}{\cal{T}}^-(x){\cal{T}}^+(1)\bar{\alpha}_{-(1-\frac{k}{n})}\mid t\rangle}
{\langle \bar{t}\mid\alpha_{1-\frac{k}{n}}\bar{\alpha}_{-(1-\frac{k}{n})}\mid t\rangle
\langle \bar{t}\mid{\cal{T}}^-(x){\cal{T}}^+(1)\mid t\rangle }
e^{-\frac{2\pi}{\beta}(1-\frac{k}{n})}+\mbox{anti-holomophic terms}+O(e^{-\frac{2\pi}{\beta}})\right\}. \notag \\
\eea
Under the conformal transformation, each term in (\ref{mod0}) transforms into a four-point function, with two twist operators and two operators from the twist sector.
As shown in \cite{Dixon0},
\be \langle \bar{t}\mid{\cal{T}}^-(x){\cal{T}}^+(1)\mid t\rangle =(x(1-x))^{-\frac{2k}{n}(1-\frac{k}{n})}
\frac{1}{F(\frac{k}{n},1-\frac{k}{n},1,x)F(\frac{k}{n},1-\frac{k}{n},1,1-x)}. \ee
The other terms in  (\ref{mod}) can be captured by the two-point functions on the orbifold.
Since in the twist sector, $\alpha_{m-\frac{k}{n}}$ and $\bar{\alpha}_{m+\frac{k}{n}}$ can be written as
\bea &~&\alpha_{m-\frac{k}{n}}=\frac{1}{2\pi i}\oint dz\partial \Phi z^{m-\frac{k}{n}}, \notag \\
&~& \bar{\alpha}_{m+\frac{k}{n}}=\frac{1}{2\pi i}\oint dz\partial\bar{\Phi}z^{m+\frac{k}{n}}, \eea
we find that
\bea \lefteqn{\frac{\langle \bar{t}\mid\bar{\alpha}_{\frac{k}{n}}{\cal{T}}^-(x){\cal{T}}^+(1)\alpha_{-\frac{k}{n}}\mid t\rangle}
{\langle \bar{t}\mid{\cal{T}}^-(x){\cal{T}}^+(1)\mid t\rangle} }\notag \\
&=&\frac{1}{2\pi i}\oint_0 dz z^{-\frac{k}{n}}\frac{1}{2\pi i}\oint_{\inf} dz' z'^{\frac{k}{n}}
\langle \partial \bar{\Phi}(z')\partial \Phi(z)\rangle\mid_{4~twists}, \\
\langle \bar{t}\mid\bar{\alpha}_{\frac{k}{n}}\alpha_{-\frac{k}{n}}\mid t\rangle
&=&\frac{1}{2\pi i}\oint_0 dz z^{-\frac{k}{n}}\frac{1}{2\pi i}\oint_{\inf} dz' z'^{\frac{k}{n}}
\langle \partial \bar{\Phi}(z')\partial \Phi(z)\rangle\mid_{2~twists}.
\eea
Here we relate these two quantities to the two-point functions. The subscript ``4 twists" means that we need to impose the twist boundary conditions at points $\{0,x,1,\inf\}$, and ``2 twists" means we only have twist conditions at points ${0,\inf}$. The  integral over $z$  goes around the origin and the integral over $z'$ goes around the infinity. In the similar way, we have
\bea  \lefteqn{\frac{\langle \bar{t}\mid\alpha_{(1-\frac{k}{n})}{\cal{T}}^-(x){\cal{T}}^+(1)\bar{\alpha}_{-(1-\frac{k}{n})}\mid t\rangle}
{\langle \bar{t}\mid{\cal{T}}^-(x){\cal{T}}^+(1)\mid t\rangle} }\notag \\
&=&\frac{1}{2\pi i}\oint_{\inf} dz z^{(1-\frac{k}{n})}\frac{1}{2\pi i}\oint_{0} dz' z'^{-(1-\frac{k}{n})}
\langle \partial \bar{\Phi}(z')\partial \Phi(z)\rangle\mid_{4~twists},
\eea
\bea \langle \bar{t}\mid\alpha_{(1-\frac{k}{n})}\bar{\alpha}_{-(1-\frac{k}{n})}\mid t\rangle
&=&\frac{1}{2\pi i}\oint_{\inf} dz z^{1-\frac{k}{n}}\frac{1}{2\pi i}\oint_{0} dz' z'^{-(1-\frac{k}{n})}
\langle \partial \bar{\Phi}(z')\partial \Phi(z)\rangle\mid_{2~twists}.
\eea
The two-point functions with 4 twists and 2 twists have been computed in \cite{Dixon0}
\be -\frac{1}{2}\langle \partial_z \Phi(z)\partial_{z'} \bar{\Phi}(z')\rangle\mid_{2~twists}=
z^{-(1-\frac{k}{n})}z'^{-\frac{k}{n}}\left(\frac{(1-\frac{k}{n})z+\frac{k}{n}z'}{(z-z')^2}\right), \ee
and
\bea  \lefteqn{-\frac{1}{2}\langle \partial_z \Phi(z)\partial_{z'} \bar{\Phi}(z')\rangle\mid_{4~twists}}\notag \\
&=& w_{n-k}(z)w_{k}(z')
\left\{(1-\frac{k}{n})\frac{z(z-1)(z'-x)}{(z-z')^2}+\frac{k}{n}\frac{(z-x)z'(z'-1)}{(z-z')^2}+A(x)\right\},
\eea
where
\bea &~&w_k(z)=[z(z-1)]^{-\frac{k}{n}}[z-x]^{-(1-\frac{k}{n})}, \notag\\
&~& w_{n-k}(z)=[z(z-1)]^{-(1-\frac{k}{n})}[z-x]^{-\frac{k}{n}}, \notag\eea
and
\be A(x)=x(1-x)\left(\frac{F^{'}(x)}{2F(x)}-\frac{F^{'}(1-x)}{2F(1-x)}\right). \ee
We have already set $(z_1,z_2,z_3,z_4)$ to $(0,x,1,\inf)$ and  $x$ to be on the real axis. With the above results, we finally obtain
\bea \frac{\langle \bar{t}\mid\bar{\alpha}_{\frac{k}{n}}{\cal{T}}^-(x){\cal{T}}^+(1)\alpha_{-\frac{k}{n}}\mid t\rangle}
{\langle \bar{t}\mid{\cal{T}}^-(x){\cal{T}}^+(1)\mid t\rangle
\langle \bar{t}\mid\bar{\alpha}_{\frac{k}{n}}\alpha_{-\frac{k}{n}}\mid t\rangle}
&=&x^{-\frac{k}{n}}(x-\frac{n}{k}A(x)), \eea
and
\bea \frac{\langle \bar{t}\mid\alpha_{(1-\frac{k}{n})}{\cal{T}}^-(x){\cal{T}}^+(1)\bar{\alpha}_{-(1-\frac{k}{n})}\mid t\rangle}
{\langle \bar{t}\mid{\cal{T}}^-(x){\cal{T}}^+(1)\mid t\rangle
\langle \bar{t}\mid\alpha_{(1-\frac{k}{n})}\bar{\alpha}_{-(1-\frac{k}{n})}\mid t\rangle}
&=&x^{-(1-\frac{k}{n})}(x-\frac{n}{n-k}A(x)). \eea
Inserting them back into (\ref{mod}) and using
 the recursion relations (\ref{recursion}), we find that the partition function (\ref{mod}) is the same as (\ref{Z}) up to orders $e^{-\frac{2\pi}{\beta}\frac{k}{n}}$ and $e^{-\frac{2\pi}{\beta}(1-\frac{k}{n})}$.
This agreement supports our prescription that in the large interval limit we should insert the complete basis from the twist sector to compute the partition function. 

\section{Thermal and Entanglement entropy}

In this section, we try to study the relation between the thermal entropy and the entanglement entropy for a 2D CFT. We  show that the relation (\ref{th}) could be proved if we insert  the  complete basis from the twist sector and use the correspondence (\ref{tw}) between the twist sector for the large interval case state and the normal sector state. The proof is very general, we only need to assume that the CFT has a discrete spectrum\footnote{The main result in this section has been briefly reported in \cite{ChenWu2}.}.

At high temperature, we quantize the theory along the spatial direction of length $R$.  Then we have the partition function
\bea Z_1&=&\int[d\phi]e^{-S_E} \notag\\
&=&\Tr e^{-RH} \notag\\
&=&e^{R\frac{2\pi}{\beta}\frac{c}{12}}\sum_ie^{-\frac{2\pi R}{\beta}\Delta_i}, \eea
where the summation is over all the excited states with conformal dimension $\Delta_i$.
For the partition function on $n$-sheeted Riemann surface $Z_n$, we can insert in a complete basis in the twist sector along the B cycle in Fig. \ref{i2}. We find  that the partition function is a sum of the four-point correlation functions
\bea\label{par} Z_n&=&e^{\frac{\pi c}{6n}\frac{R}{\beta}}\sum_i\langle t,i\mid {\cal{T}}^+(u_1){\cal{T}}^-(u_2)\mid t,i\rangle
e^{-\frac{2\pi R}{n\beta}\Delta_i}, \eea
where the summation is over all the states in the twisted sector. We have used the fact that the conformal dimension of the twist  sector state is related to the one of the normal sector state by the relation (\ref{tw}).
For the large interval, the complement part of the interval is small and we can use the OPE of  two twist operators
\be {\cal{T}}^{+}(u+\frac{l}{2}){\cal{T}}^{-}(u-\frac{l}{2})\sim c_nl^{-\frac{c}{6}(n-\frac{1}{n})}(1+O(l)), \ee
from which we have
\bea S_n&=&-\frac{1}{n-1}(\log~Z_n-n\log~Z_1) \notag \\
&=&-\frac{1}{n-1}\left((\frac{\pi c}{6n}\frac{R}{\beta}+\log~(c_nl^{-\frac{c}{6}(n-\frac{1}{n})})
+\log(\sum_ie^{-\frac{2\pi R}{n\beta}\Delta_i})+O(l))\right. \notag \\
&~&\left. -n(\frac{\pi c}{6}\frac{R}{\beta}+\log(\sum_ie^{-\frac{2\pi R}{\beta}}))\right),
\eea
and the entanglement entropy
\be S_{EE}=\lim_{n\rightarrow 1}S_n. \ee
The quantity we are interested in is
\bea\label{th1} \lefteqn{\lim_{l\rightarrow0}(S_{EE}(R-l)-S_{EE}(l))}\notag \\
&=&\frac{\pi c}{3}\frac{R}{\beta}-\lim_{n\rightarrow1}\frac{1}{n-1}(\log~Z[\frac{R}{n\beta}]-n\log~Z[\frac{R}{\beta}]) \notag \\
&=&\frac{\pi c}{3}\frac{R}{\beta}+\log Z[\frac{R}{\beta}]+\frac{R}{\beta}\frac{Z^{'}[\frac{R}{\beta}]}{Z[\frac{R}{\beta}]},
\eea
with
\be Z[x]=\sum_ie^{-2\pi x\Delta_i}. \ee
Note that the terms proportional to $c_n$ have been canceled.

On the other hand,  the partition function at high temperature  could be read by a modular transformation
\be Z_{CFT}=e^{\frac{c\pi}{6}\frac{R}{\beta}}Z[\frac{R}{\beta}]. \ee
And the thermal entropy could be obtained by
\bea\label{spacial}\lefteqn{S_{th}=-\frac{\partial F}{\partial T}=-\beta^2\frac{\partial}{\partial\beta}(\frac{1}{\beta}\log~Z_{CFT})} \notag \\
&=&\frac{1}{3}\pi c\frac{R}{\beta}-(\beta^2\frac{\partial}{\partial\beta}\frac{1}{\beta}\log~Z[\frac{R}{\beta}]) \notag \\
&=&\frac{1}{3}\pi c\frac{R}{\beta}+\log Z[\frac{R}{\beta}]+\frac{R}{\beta}\frac{Z^{'}[\frac{R}{\beta}]}{Z[\frac{R}{\beta}]},
\eea
which is the same as (\ref{th1}). Therefore we prove the relation (\ref{th}) for a general CFT. The key point in our proof is the insertion of the complete basis of the twist sector and the one-to-one correspondence between the twist sector state and the normal sector state.

 Note that our proof for (\ref{th}) does not rely on the temperature. Even for the low temperature, our proof still works, though the relation (\ref{th1}) is reminiscent of the high temperature result for a CFT with holographic dual. As for a CFT its thermal partition function is modular invariant, we can always insert a complete basis at a spacial cycle or a thermal one. In the above discussion, we quantized the theory along the spatial direction and inserted the basis at the thermal cycle. Instead, we can also quantize the theory along the thermal direction and insert a  set of complete basis from normal sector at the spacial cycle.  The partition function now becomes
\be Z_1=e^{\frac{\pi c}{6}\frac{\beta}{R}}\sum_ie^{-2\pi\frac{\beta}{R}\Delta_i}
=e^{\frac{\pi c}{6}\frac{\beta}{R}}Z[\frac{\beta}{R}]. \ee
For small interval,
\be {\cal{T}}^+(-\frac{l}{2}){\cal{T}}^-(\frac{l}{2})\sim c_nl^{-\frac{c}{6}n(1-\frac{1}{n^2})}(1+O(l)), \ee
and
\bea Z_n&=&\sum_{i_1,i_2...,i_n}(\prod_{k=1}^{n}e^{\frac{\pi c}{6}\frac{\beta}{R}}
e^{-2\pi\frac{\beta}{R}\Delta_{i_k}})
\langle i_1,i_2,...i_n\mid{\cal{T}}^+(-\frac{l}{2}){\cal{T}}^-(\frac{l}{2})\mid i_1,i_2,...i_n\rangle \notag \\
&=&c_nl^{-\frac{c}{6}n(1-\frac{1}{n^2})}e^{\frac{\pi c}{6}\frac{n\beta}{R}}(Z[\frac{\beta}{R}]^n+O(l)),
\eea
so
\be S_n=-\frac{1}{n-1}\log~\frac{Z_n}{Z_1^n}=-\frac{1}{n-1}\log~(c_nl^{-\frac{c}{6}n(1-\frac{1}{n^2})})+O(l). \ee
For large interval, to calculate $Z_n$ we may also insert a complete basis at spacial cycles. However, at this time, we need to clarify the states in different sheets. In Fig. \ref{i2},  the upper side and lower side are in  the same sheet if they are connected by  the dotted line. However in large interval limit, the dotted line  shrinks to zero size in the leading order of the OPE of twist operators.  In this case the interval effectively crosses over the whole spatial direction, so the state at the upper side and the state at the low side are in different replicas. More precisely, the state at the upper side of the $i$-th replica changes to the state at the low side of the $(i+1)$-th replica. Therefore we need to replace the states $\mid i_1,i_2,...i_n\rangle$ in the correlation function to $\mid i_2,i_3,...i_n,i_1\rangle$.  So for the large interval the partition function is
\bea Z_n&=&\sum_{i_1,i_2...,i_n}(\prod_{k=1}^{n}e^{\frac{\pi c}{6}\frac{\beta}{R}}
e^{-2\pi\frac{\beta}{R}\Delta_{i_k}})
\langle i_1,i_2,...i_n\mid{\cal{T}}^+(-\frac{l}{2}){\cal{T}}^-(\frac{l}{2})\mid i_2,i_3,...i_n,i_1\rangle \notag \\
&=&c_nl^{-\frac{c}{6}n(1-\frac{1}{n^2})}e^{\frac{\pi c}{6}\frac{n\beta}{R}}(Z[\frac{n\beta}{R}]+O(l)).
\eea
Namely, only the states identified in all the replicas contribute to the leading order partition function. Consequently
\be S_n=-\frac{1}{n-1}\log~\frac{Z_n}{Z_1^n}=-\frac{1}{n-1}\big(\log~(c_nl^{-\frac{c}{6}n(1-\frac{1}{n^2})})
+\log~\frac{Z[\frac{\beta}{R}]^n}{Z[\frac{n\beta}{R}]}\big)+O(l), \ee
and
\be \lim_{\epsilon\rightarrow0}S_{EE}(R-\epsilon)-S_{EE}(\epsilon)=\log~Z[\frac{\beta}{R}]-
\frac{\beta}{R}\frac{Z^{'}[\frac{\beta}{R}]}{Z[\frac{\beta}{R}]}.\label{th2}\ee
Similarly, for the thermal entropy,
\be Z_{CFT}=e^{\frac{c\pi}{6}\frac{\beta}{R}}Z[\frac{\beta}{R}], \ee
and
\be\label{time} S_{th}=-\frac{\partial F}{\partial T}=\log Z[\frac{\beta}{R}]-\frac{\beta}{R}\frac{Z^{'}[\frac{\beta}{R}]}{Z[\frac{\beta}{R}]} \ee
which is just the relation (\ref{th2}).

We have already proved (\ref{th}) from the points of view of both spacial direction quantization and time direction quantization. The relations (\ref{spacial}) and (\ref{time}) equals each other due to  modular transformation. Both of them can be written as the summation over infinite series. For the high temperature the series from  (\ref{spacial}) converge faster, while for the low temperature the series from (\ref{time}) converge faster. This is especially clear for the large central charge limit of the CFT holographically corresponding to pure AdS$_3$ quantum gravity\cite{small}.
 For low temperature,  the leading order in the large $c$ expansion has only the $c^0$ term from the expansion (\ref{time}). This is consistent with the fact that in the holographic description the dual spacetime is a thermal AdS and there is no difference between $S_{EE}(R-\epsilon)$ and $S_{EE}(\epsilon)$ at classical order. For the
  high temperature  there is a term of order $c^1$ from (\ref{spacial}), which is just the entropy of the BTZ black hole. In the holographic description, the bulk spacetime is a BTZ black hole so the entanglement entropy of the very large interval is the horizon length plus the geodesic connecting the complement interval, and therefore $S_{EE}(R-\epsilon)-S_{EE}(\epsilon)$ reproduces exactly the black hole entropy\cite{Takayanagi}. The remaining terms in (\ref{th1}) are the quantum corrections to the black hole entropy.



In the above proof we assume that the CFT has a discrete spectrum. For a CFT with continuous spectrum, the situation is not clear. Actually  the non-compact boson presents  an interesting example. This could be seen directly from the expansion of the $W$ functions (\ref{Z}). Asymptotically as $z \to 1$ the hypergeometric function behaves
\be F[\frac{k}{n},1-\frac{k}{n},1,z]\sim-\frac{\sin(\frac{\pi k}{n})}{\pi}\log(1-z). \ee
So the leading term in the R\'enyi entropy is
\bea S_n(R-\epsilon)&=&-\frac{1}{n-1}\log(\prod_{k=1}^{n-1}\frac{Z_{k,n}}{Z_1}) \notag \\
&=&-\frac{1}{n-1}\log (\tilde{c}_n\epsilon^{-\frac{1}{3n}(n^2-1)})-\frac{1}{n-1}\log(\log(\frac{2\pi \epsilon}{\beta})^{-(n-1)})+\mbox{finite terms} \notag \\
&=&-\frac{1}{n-1}\log \tilde{c}_n+\frac{n+1}{3n}\log~\epsilon+\log(|\log\epsilon|)+\mbox{finite terms}, \eea
where the log-logarithmic term $\log(|\log\epsilon|)$ is a new kind of divergence in the large interval limit. This log-logarithmic term appear also in the small regime of double interval R\'enyi entropy for free non-compact boson \cite{Cardy1}. It originates from the continuous spectrum of the theory.
It could not be cancelled by any term in the short interval expansion of the R\'enyi entropy. Nevertheless, we notice that after removing this divergence  the relation (\ref{th}) still holds up to a constant.

We can evaluate (\ref{partition}) in both the small interval and large interval limits. For the small interval, we can use the relation (\ref{sW}) and
\be \vartheta_1(u_2-u_1\mid \tau)=\vartheta_1^{'}(0\mid \tau)(u_2-u_1)+O((u_2-u_1)^2), \ee
to get
\be \frac{Z_n}{Z_1^n}=|u_2-u_1|^{-\frac{1}{3}(n-\frac{1}{n})}(1+O(u_2-u_1)). \ee
For the large interval, we use (\ref{lW1}, \ref{lW2}) and the periodicity of the theta functions to read
\bea \frac{Z_n}{Z_1^n}
&=&
|1-u_2+u_1|^{-\frac{1}{3}(n-\frac{1}{n})} \left(
\prod_{k=1}^{n-1} \left( 2\pi \beta\eta(i\beta)^{6}
\vartheta_1(-\frac{k}{n}\mid i\beta)^{-1} \vartheta_1(-(1-\frac{k}{n})\mid i\beta)^{-1} (\log|1-u_2+u_1|)^{-1} \right) \right. \notag\\
&~&+O((1-u_2+u_1)^0)
\left. \right) \notag \\
&=&|1-u_2+u_1|^{-\frac{1}{3}(n-\frac{1}{n})} \left(\beta^{n-1}(2\pi)^{n-1}
\frac{\eta(i\beta)^{4n}}{n^2\eta(in\beta)^4} \left| \log|1-u_2+u_1| \right|^{-(n-1)}+O((1-u_2+u_1)^0) \right)
,\notag \\
\eea
For the second line we have used the relation \cite{Chen:2015cna}
\be \prod_{k=1}^{n-1} \left( \eta(i\beta)^{-3}\vartheta_1(-\frac{k}{n}\mid i\beta) \right)=n
\frac{\eta(in\beta)^{2}}{\eta(i\beta)^{2n}}. \ee
With this relation, we find
\bea \label{nonRen}
\lefteqn{\lim_{l\rightarrow 0} S_n(1-l)-S_n(l)} \notag \\
&=&-\frac{1}{n-1}\left[(n-1)\log(2\pi\beta)-2\log n+4n\log\eta(i\beta)-4\log\eta(in\beta)
-(n-1)\log(\left|\log~|l|\right|)\right], \notag \\
\eea
and for the entanglement entropy
\bea
\lefteqn{\lim_{l\rightarrow 0} S_{EE}(1-l)-S_{EE}(l)} \notag \\
&=&-\log2\pi\beta+2-4\log\eta(i\beta)+4i\beta\frac{\eta^{'}(in\beta)}{\eta(in\beta)}+\log|\log~l|.
\eea
On the other hand, considering the partition function
\bea Z&=&\left[\frac{V}{2\pi}\int dk\sum_{\{m\},\{\bar{m}\}} e^{-2\pi\beta(k^2+m+\bar{m}-\frac{1}{12})} \right]^2
 \notag \\
&=&(\frac{V}{2\pi})^2\frac{1}{2\beta}\frac{1}{\eta(i\beta)^4} ,
\eea
and (\ref{time}), we get the thermal entropy
\be S_{th}=\frac{4i\beta\eta^{'}(i\beta)}{\eta(i\beta)}-4\log \eta(i\beta)-\log \beta
+\log\frac{V^2}{8\pi^2}+1. \ee
The relation between thermal and entanglement entropy is modified to be
\bea \lim_{l\rightarrow 0}
S_{EE}(1-l)-S_{EE}(l)-\log|\log l|
&=&S_{th}(\beta)+c_1, \eea
where
\be c_1=1+\log\frac{4\pi}{V^2}, \ee
and $V$ is the volume of the target space. In the above discussion, we have taken the point of view that the non-compact boson is the large volume limit of the compact boson, where all of the physical dimensionless observable is larger than $\frac{1}{V}$. The order of taking the limits is essential. We will go back to this problem later.

We note that the log-logarithmic divergence comes from the continuous spectrum of the non-compact free boson. We may use the OPE of the twist operators  to reproduce the result (\ref{nonRen}). In this part, let us still consider a real scalar for simplicity.
Before the calculation, we need to clarify the regularization of the continuous spectrum. We regard the non-compact boson as a large volume limit of a compact boson, therefore  we have
\be\label{regulation} \sum_k \rightarrow \frac{V}{2\pi}\int dk,~~~ \delta_{k_1,k_2} \rightarrow \frac{2\pi}{V}\delta(k_1-k_2),
~~~\delta(0)=\frac{V}{2\pi}. \ee
   In the OPE of two twist operator we have
\bea\label{OPEc}  {\cal{T}}^+(-\frac{l}{2}){\cal{T}}^-(\frac{l}{2})
&=& c_nl^{-\frac{1}{6}n(1-\frac{1}{n^2})}\big[(\frac{V}{2\pi})^{n-1}\int\prod_{j=1}^ndk_j\delta(\sum_{j=1}^n k_j)
\prod_{j=1}^n(\frac{l}{n})^{k_j^2}e^{-ik_jX^{(j)}(u)}\mid_{u=0} \notag \\
&~&\cdot \prod_{1\leq j_1<j_2\leq n}(2\sin\frac{\pi}{n}(j_2-j_1))^{2k_{j_1}k_{j_2}}+O(l)\big],
\eea
where V is the regularized volume of target space.
We follow the second proof with time direction quantization, and insert a complete basis along  spacial cycles. For the small interval, we find
\bea Z_n
&=&
 c_nl^{-\frac{1}{6}n(1-\frac{1}{n^2})}\big[(\frac{V}{2\pi})^{n-1}\int\prod_{j=1}^ndk_j\delta(\sum_{j=1}^n k_j)
\prod_{j=1}^n(\frac{l}{n})^{k_j^2}
\int \frac{V}{2\pi}dp_j  \notag \\
&~&\sum_{\{m_j\},\{\bar{m_j}\}} e^{-2\pi\beta(p_j^2+m_j+\bar{m_j})}\langle p_j,\{m_j\},\{\bar{m_j}\} |e^{-ik_jX^{(j)}(u)} |p_j,\{m_j\},\{\bar{m_j}\}\rangle\mid_{u=0} \notag \\
&~&\cdot \prod_{1\leq j_1<j_2\leq n}(2\sin\frac{\pi}{n}(j_2-j_1))^{2k_{j_1}k_{j_2}}+O(l)\big] \notag \\
&=&c_nl^{-\frac{1}{6}n(1-\frac{1}{n^2})}\big[(\frac{V}{2\pi})^{n-1}\int\prod_{j=1}^ndk_j\delta(\sum_{j=1}^n k_j)
\prod_{j=1}^n(\frac{l}{n})^{k_j^2}
\int \frac{V}{2\pi}dp_j \notag \\
&~&\sum_{\{m_j\},\{\bar{m_j}\}} e^{-2\pi\beta(p_j^2+m_j+\bar{m_j})}  \frac{2\pi}{V}\delta(k_j)
 \prod_{1\leq j_1<j_2\leq n}(2\sin\frac{\pi}{n}(j_2-j_1))^{2k_{j_1}k_{j_2}}+O(l)\big] \notag \\
&=&c_nl^{-\frac{1}{6}n(1-\frac{1}{n^2})}((\frac{V}{2\pi})^n(\frac{1}{2\beta})^{\frac{n}{2}}\eta(i\beta)^{-2n}
+O(l)),
\eea
where $\{m_j\}$ $\{\bar{m_j}\}$ denote the excitation on the primary states, and for the second equation we have used the momentum conservation.

For the large interval, the situation is different. The non-zero momentum operators $\prod_{j=1}^ne^{ik_jX^{(j)}}$ can have a contribution to the partition function in the large interval limit. Taking (\ref{OPEc}) into the partition function, the partition function transforms into a summation of $n$-point correlation on a torus. To the leading order, we obtain
\bea Z_n
&=&c_nl^{-\frac{1}{6}n(1-\frac{1}{n^2})}\big[(\frac{V}{2\pi})^{n-1}\prod_{j=1}^n \int dk_j
\delta(\sum_{j=1}^nk_j)\prod_{j=1}^n(\frac{l}{n})^{k_j^2}
\prod_{1\leq j_1<j_2\leq n}(2\sin\frac{\pi}{n}(j_2-j_1))^{2k_{j_1}k_{j_2}} \notag \\
&~&\cdot \langle \prod_{j=1}^n e^{-ik_jX(u_j)} \rangle\mid _{n\beta~\mbox{\tiny torus}} +O(l)\big] \notag \\
&=&c_n l^{-\frac{1}{6}n(1-\frac{1}{n^2})}[(\frac{V}{2\pi})^{n-1} \prod_{j=1}^n \int dk_j
\delta(\sum_{j=1}^nk_j)\prod_{j=1}^n(\frac{l}{n})^{k_j^2}\notag
\prod_{1\leq j_1<j_2\leq n}(2\sin\frac{\pi}{n}(j_2-j_1))^{2k_{j_1}k_{j_2}} \notag \\
&~&\cdot \frac{V}{2\pi}\int dp \sum_{\{m\},\{\bar{m}\}}
e^{-2\pi n\beta(p^2+m+\bar{m}-\frac{1}{12})}
\langle p,m\mid \prod_{j=1}^n e^{-ik_jX(u_j)} \mid p,m\rangle
\mid_u+O(l)],\notag \\
\eea
where $u_j=i(j-1)\beta $. In the large interval limit, $l$ is close to zero, only small $k$ gives a dominant contribution. We can set all of the momentum except the one on the power of $l$ to zero. Then we read
\bea Z_n&=&c_nl^{-\frac{1}{6}n(1-\frac{1}{n^2})}((\frac{V}{2\pi})^n
\frac{1}{(2n\beta)^{\frac{1}{2}}}
\eta(in\beta)^2
\prod_{j=1}^n\int dk_j \delta(\sum_{j=1}^n k_j)\prod_{j=1}^n l^{k_j^2}+O(l^0)) \notag \\
&=&c_nl^{-\frac{1}{6}n(1-\frac{1}{n^2})}((\frac{V}{2\pi})^n
\frac{1}{(2n\beta)^{\frac{1}{2}}}
\eta(in\beta)^{-2}
\cdot \frac{\pi^{\frac{n-1}{2}}}{n^{\frac{1}{2}}} |\log l|^{-\frac{n-1}{2}}+O(l^0)).
 \eea
 Finally we obtain
 \bea \lefteqn{\lim_{l\rightarrow 0}S_n(1-l)-S_n(l)} \notag \\
 &=&\lim_{l\rightarrow 0}-\frac{1}{n-1}(Z_n(1-l)-Z_n(l)) \notag \\
 &=&-\frac{1}{n-1}\big[\frac{n-1}{2}\log 2\pi \beta-\log n+
 2n\log \eta(i\beta)-2\log \eta(in\beta) -\frac{n-1}{2}\log |\log l| \big], \notag
 \eea
 which is exactly half of (\ref{nonRen}), as we are considering a real scalar.
 
\begin{figure}[tbp]
  \centering
\includegraphics[width=8cm]{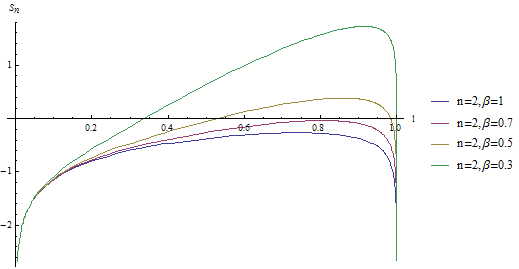}

  \caption{The picture for $S_n$ for $n=2$ and $\beta =0.3,0.5,0.7,1$. It is divergent near $l \to 0$ and $l\to 1$, but the divergences are different.}  \label{Sn1}     \end{figure}
  
 \begin{figure}[tbp]
  \centering
 \includegraphics[width=8cm]{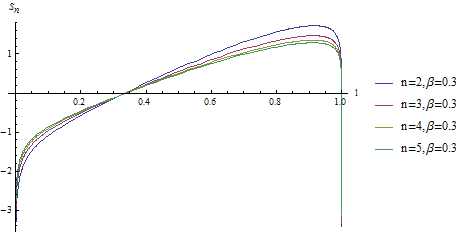}
 
   \caption{The picture for $S_n$. For $n=2,3,4,5$ $\beta=0.3$.} \label{Sn2}  \end{figure} 
   
     \begin{figure}[tbp]
  \centering
   \subfloat[$S_n-\frac{n+1}{3n}\log l$] {\includegraphics[width=6cm]{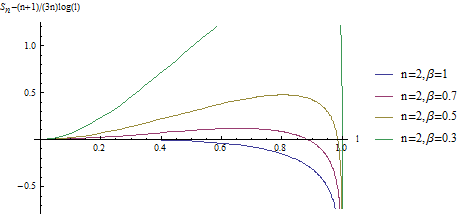}
  \label{Sn10}}
  \subfloat[$S_n-\frac{n+1}{3n}\log(1-l)-\log|\log(1-l)|$ ] {\includegraphics[width=6cm]{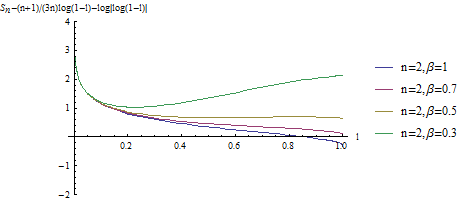} 
  \label{Sn11}} \\
  \caption{ The figures for the entropies with divergences being subtracted at one end. These figures confirm the divergent behaviors at $l \to 0$ and $l \to 1$. }\label{S1}
  \end{figure}

   \begin{figure}[tbp]
  \centering 
   \subfloat[$S_n-\frac{n+1}{3n}\log l$] {\includegraphics[width=6cm]{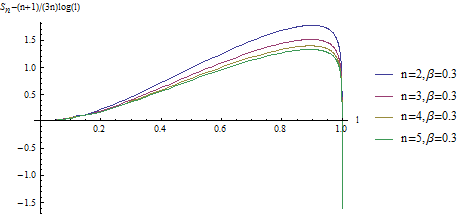}
  \label{Sn20}}
  \subfloat[$S_n-\frac{n+1}{3n}\log(1-l)-\log|\log(1-l)|$ ] {\includegraphics[width=6cm]{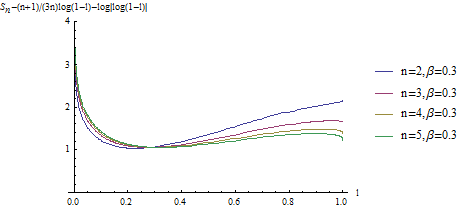}
  \label{Sn21}} \\
  \caption{The picture for $S_n$ after the divergence being subtracted at one end. For $n=2,3,4,5$ and $\beta=0.3$.}
  \end{figure}

To support the analytic result, we do some numerical calculations for the R\'enyi entropies. In Fig. \ref{Sn1}, we plot $S_2$ for different temperatures. In Fig. \ref{Sn2}, we plot $S_n, n=2,3,4,5$ for a fixed $\beta=0.3$. The divergences in the small interval and large interval limits are different. For the small interval, 
\be S_n=\frac{n+1}{3n}\log l +O(l) \ee
for the large interval 
\be S_n=\frac{n+1}{3n}\log (1-l)+\log|\log(1-l)|+O((1-l)^0) \ee
To check these divergent behaviors, we subtract such divergences from $S_n$ and plot the subtracted entropies in Fig. \ref{Sn10}, \ref{Sn11} and Fig.\ref{Sn20}, \ref{Sn21}. After subtraction, the remaining entropies are indeed finite. 

The presence of log-logarithmic terms in the above discussion is remarkable. As we have shown and would like to emphasize, it is due to the continuous spectrum of the theory.
If we regularize the theory by considering a compact free scalar such that the spectrum becomes discrete and the theory is gapped, the relation (\ref{th}) indeed holds, as checked explicitly in \cite{Chen:2015cna}. Actually here is a subtle order-of-limits issue. We may
regard the noncompact scalar as the large volume limit of compact scalar, namely $V\to \infty$. However, this limit is not commutative with the large interval limit $l\to 0$. In fact, in the OPE (\ref{OPEc}) of the twist operators the first primary field is $e^{\frac{i}{V}X(u)}$ and the coefficient for this term is proper to $l^{-\frac{1}{V^2}}$. This term goes to zero in the $l\rightarrow 0$ limit, while it goes to $1$ in the continuous spectrum limit when $V\rightarrow \inf$. It is clear these two limits do not commute. In the above discussion, we have taken the point of view that the $V\rightarrow \inf$ limit should be taken first and then the $l\rightarrow 0$ limit, this leads to the above log-logarithmic divergence. On the contrary, if we take the $l\rightarrow 0$ limit first and finally take the $V\rightarrow \inf$ limit, we reproduce the relation (\ref{th}). In this order, we actually consider a compact scalar and take the noncompact limit at the end.



Another remarkable fact on the log-logarithmic term is that it may exist at the zero temperature limit if we insist on the noncompact scalar without regularization. Here it involves another subtle order-of-limits issue. Let us always take the $V\rightarrow \inf$ limit first to have a theory with continuous spectrum. As we know, at zero temperature, the entanglement entropy is of a universal form and there is always $S_{EE}(R-l)-S_{EE}(l)=0$ so that the relation (\ref{th}) holds. However, we also notice that the zero temperature limit $T\to 0$ and the large interval limit $l\to 0$ is not commutative. If we instead take the $l \to 0$ limit first and then the $T\to 0$ limit next, we would find the log-logarithmic divergence again. Generally for a theory with degenerate vacuum or with a continuous spectrum, the order in taking the limit is important. Different orders  may give different results, as already shown in \cite{Herzog:2013py,Cardy2}. This happens in the case at hand as well.


\section{Conclusion and discussion}

In this paper, we studied the large interval R\'enyi entropy of 2D CFT at high temperature. 
We proposed that in the large interval limit we should insert a complete twist sector states along the $B$ cycle in Fig. \ref{i2} to compute the partition function more effectively. We discussed the twist sector in a general CFT module and showed that there is a one-to-one correspondence between the twist sector states and the normal sector states.

To check our prescription for computing the R\'enyi entropy, we revisited the non-compact free boson theory. We corrected the treatments in \cite{Datta} and expanded the $W$ functions explicitly in both the short interval and large interval limits. In the small interval limit, besides using the expansion of $W$ functions, we computed the partition functions in  two other ways. First we used the OPE of the twist operators in the limit and found agreement with direct expansion. This supports our treatment on the $W$ functions. The other way is to insert the complete normal sector basis to compute the correlation functions.  In the large interval limit, using the proposal to insert the complete twist sector states we calculated the first few leading terms of the R\'enyi entropy. We found good agreement with the result obtained from large interval expansion of $W$ functions. This strongly supports our proposal to compute the large interval limit of R\'enyi entropy at high temperature.

We discussed the relation between the thermal entropy and the entanglement entropy. We proved the relation (\ref{th}) at any temperature using two different approaches for the CFT with a discrete spectrum. The rational CFT belongs to this class. We note that the proof based on the time direction quantization seems to be more general. It  does not depend on the modular invariance of the theory. For example, the free Dirac fermion with antiperiodic  boundary condition has been discussed in \cite{Takayanagi}, where the relation (\ref{th}) has been proved. However, the free fermion with only NS sector is not modular invariant. Nevertheless we can prove the relation via time direction quantization formalism. Moreover, we also showed that for non-compact free scalar, the relation
breaks down, as there appears a log-logarithmic term in the large interval limit of the R\'enyi entropy. Such a divergent term originates from the continuous spectrum of the theory. However, after properly regularizing the theory,  the relation (\ref{th}) is recovered. It would be interesting to study this relation for a generic 2D quantum field theory or generalize it to higher dimension\cite{Hubeny:2013gta}.

Our proposal could be applied to the study of other CFTs. For example, it would be interesting to generalize the study to the compact free scalar. For the compact case, the R\'enyi entropy has been given in terms of theta functions\cite{Datta}. In this case, the zero modes include both the momenta and the winding modes.  The first excitation may have conformal dimension $(\frac{V^2}{4\alpha'},\frac{V^2}{4\alpha'})$ for small radius or $(\frac{\alpha'}{4V^2},\frac{\alpha'}{4V^2})$ for a large radius. And the leading thermal correction at low temperature should be
\be \frac{1}{n-1}(n-\frac{1}{n^{2\Delta-1}}
\frac{\sin^{2\Delta}(\pi l)}{\sin^{2\Delta}(\frac{\pi l}{n})})e^{-2\pi \Delta\frac{\beta}{R}}, \ee
so we need to recalculate the R\'enyi entropy in this model.  


In the short interval case, there is a universal thermal correction for a primary field \cite{Cardy2}. However such universal behavior is absent in the large interval limit via our approach. Suppose that there is a light primary operator with conformal dimension $(h,\bar{h})$ in the theory, after some calculation we find that
\bea Z_n=e^{\frac{\pi c}{6n\beta}}\big \langle t\mid {\cal{T}}^-(-\frac{l}{2}){\cal{T}}^+(\frac{l}{2})\mid t\rangle
+e^{-\frac{2\pi}{n\beta}(h+\bar{h})}
\langle t,\phi\mid {\cal{T}}^-(-\frac{l}{2}){\cal{T}}^+(\frac{l}{2})\mid t,\phi\rangle. \label{univ} \eea
Let us first consider the first correlator in (\ref{univ})
\bea \label{primary} &~&\langle t\mid{\cal{T}}^-(-\frac{l}{2}){\cal{T}}^(\frac{l}{2})\mid t\rangle \notag \\
&=&c_nl^{-\frac{1}{6}cn(1-\frac{1}{n^2})}
\left(1+(-1)^{2(h-\bar{h})}(\frac{1}{2n})^{4(h+\bar{h})}(\frac{2\pi}{\beta})^{2(h+\bar{h})}\frac{n}{2}
\sum_{j=1}^{n-1}\frac{1}{(\sin\frac{\pi}{n}j)^{4(h+\bar{h})}}l^{2(h+\bar{h})}+...\right). \notag \\
\eea
In the OPE of the twist operators, there are contributions from the primary fields. Such correction really takes a universal form. For the second correlator in (\ref{univ}), we obtain
\bea &~&\langle t,\phi\mid {\cal{T}}^-(-\frac{l}{2}){\cal{T}}^+(\frac{l}{2})\mid t,\phi\rangle \notag \\
&=&c_nl^{-\frac{1}{6}cn(1-\frac{1}{n^2})}
\left(1+(-1)^{2(h-\bar{h})}(\frac{1}{2n})^{4(h+\bar{h})}(\frac{2\pi}{\beta})^{2(h+\bar{h})}\frac{n}{2}
\sum_{j=1}^{n-1}\frac{F(e^{\frac{2\pi i}{n}(j_1-j_2)})}{(\sin\frac{\pi}{n}j)^{4(h+\bar{h})}}l^{2(h+\bar{h})}+...\right), \notag \\
\eea
where
\be F(x)=\lim_{z_4\rightarrow \inf}z_4^{2h}\bar{z}_4^{2\bar{h}}
\langle \phi(0)\phi(x)\phi(1)\phi(z_4) \rangle \ee
is a four-point function depending on the theory. It is clear that the leading correction is at the same order as the one in (\ref{primary}). Therefore, there is no universal correction. On the other hand, when the interval is not very large, there does exist a universal thermal correction which could be read from a modular transformation on the result in \cite{Cardy2}.


Our proposal is inspired by the holographic computation of the entanglement entropy. It would be interesting to address the issue in the context of AdS$_3$/CFT$_2$ correspondence. From the AdS$_3$/CFT$_2$ correspondence, the bulk gravitational configuration is the classical solution whose asymptotic boundary is exactly the Riemann surface  in CFT\cite{Headrick:2010zt}. In \cite{Hartman:2013mia,Faulkner:2013yia}, it has been proved that for multiple intervals in two-dimensional(2D) CFT, the leading contributions of their R\'enyi entropies in the large central charge limit are given by the classical actions of corresponding gravitational configurations. Moreover, the 1-loop quantum correction\cite{Headrick:2010zt,Barrella:2013wja} to the gravitational configuration  gives the next-leading order contribution to the R\'enyi entropy.
In the large central charge limit,  the vacuum module dominates the partition function. One may wonder if it is enough to consider only the vacuum module as the genus-$1$ thermal partition function of CFT with only vacuum module is not modular invariant\cite{Witten}. However, when we calculate the partition function, we can always cut and insert a complete basis along some cycles. When we expand the partition function in this way, the states of large conformal dimension decay very fast, and in the large $c$ limit, we can only consider the vacuum module\cite{Hartman:2013mia,Hartman:2014oaa}. This has been supported from the study of short interval R\'enyi entropy at high temperature. In this case, the holographic result is in good agreement with the field theory computation\cite{small}. In the large interval case, the situation is more subtle. On the field theory side, we need to study the twist sector of the vacuum module more carefully. On the bulk gravity side, we need to consider a different monodromy condition  from the one suggested in \cite{Barrella:2013wja}. We find good agreement in this context as well\cite{ChenWu}. More importantly, the summation series in the expansion turn out to converge nicely.


One lesson from the investigation in this paper is that in the large interval limit and at high temperature, we may have to choose different monodromy to compute the holographic R\'enyi entropy. This suggests that the similar treatment  may apply to the study of higher spin entanglement entropy\cite{Ammon1310,Boer1306,Datta1406,Datta1405,Long:2014oxa,Castro:2014mza} and the entanglement negativity\cite{Calabrese:2014yza} in the large interval limit and at high temperature.

Another interesting question is on the holographic duality of noncompact scalar discussed in \cite{Datta}. It is remarkable that the 1-loop correction in the gravity is in precise agreement with the field theory computation. Considering the discrepancy between our results and the ones in \cite{Datta}, it would be interesting to reconsider this agreement by introducing appropriate regularization in our framework\footnote{We would like to thank the anonymous referee for pointing out this possibility. }.

\vspace*{10mm}
\noindent {\large{\bf Acknowledgments}}\\
The work was in part supported by NSFC Grants No.~11275010, No.~11335012 and No.~11325522.
We would like to thank Justin David for the discussion and comments.

\vspace*{5mm}


\begin{appendix}

\section{The $W$ functions}


The functions $W_1^1$ and $W_2^2$ are defined respectively as
\bea\label{W11} W_1^1&=&\int_0^1du\vartheta_1(u-u_1\mid i\beta)^{-(1-\frac{k}{n})}\vartheta_1(u-u_2\mid i\beta)^{-\frac{k}{n}}
\vartheta_1(u-(1-\frac{k}{n})u_1-\frac{k}{n}u_2\mid i\beta) \notag \\
&=&\int_0^1due^{\frac{\pi}{\beta}\frac{k}{n}(1-\frac{k}{n})(u_1-u_2)^2}
\vartheta_1\left(\frac{u-u_1}{i\beta}\mid\frac{i}{\beta}\right)^{-(1-\frac{k}{n})}
\vartheta_1\left(\frac{u-u_2}{i\beta}\mid\frac{i}{\beta}\right)^{-\frac{k}{n}} \notag \\
&~&\cdot\vartheta_1\left(\frac{u-(1-\frac{k}{n})u_1-\frac{k}{n}u_2}{i\beta}\mid\frac{i}{\beta}\right), \eea
and
\bea\label{W22} (W_2^2)^*&=&\int_0^{i\beta}du\vartheta_1(u-u_1\mid i\beta)^{-\frac{k}{n}}\vartheta_1(u-u_2\mid i\beta)^{-(1-\frac{k}{n})}
\vartheta_1(u-\frac{k}{n}u_1-(1-\frac{k}{n})u_2\mid i\beta) \notag \\
&=&\int_0^{i\beta}due^{\frac{\pi}{\beta}\frac{k}{n}(1-\frac{k}{n})(u_1-u_2)^2}
\vartheta_1\left(\frac{u-u_1}{i\beta}\mid\frac{i}{\beta}\right)^{-\frac{k}{n}}
\vartheta_1\left(\frac{u-u_2}{i\beta}\mid\frac{i}{\beta}\right)^{-(1-\frac{k}{n})} \notag \\
&~&\cdot\vartheta_1\left(\frac{u-\frac{k}{n}u_1-(1-\frac{k}{n})u_2}{i\beta}\mid\frac{i}{\beta}\right). \eea
Here the theta function is defined by
\be \vartheta_1(u\mid\tau)\equiv2e^{\frac{1}{4}\pi i\tau}\sin \pi u\prod_{m=1}^{\inf}(1-q^m)(1-e^{2\pi iu}q^m)(1-e^{-2\pi iu}q^m), \ee
where
\be q=e^{2\pi i\tau} \ee
and in order to discuss the high temperature expansion,  we already use the $S$-duality property for theta function\cite{Polchinski}
\bea\label{Sdual} \vartheta_1(\frac{u}{\tau}\mid-\frac{1}{\tau})&=&-i(-i\tau)^{\frac{1}{2}}e^{\frac{\pi iu^2}{\tau}}\vartheta_1(u\mid\tau), \notag \\
\vartheta_1^{'}(0\mid-\frac{1}{\tau})\frac{1}{\tau}&=&-i(-i\tau)^{\frac{1}{2}}\vartheta_1^{'}(0\mid\tau). \eea

In \cite{Datta}, it was suggested that the $W$ functions are related by
\be
W^2_2=\tau W^1_1=i\b W^1_1. \label{relationW}
\ee
This relation is not true, as shown by the direct expansions in the short interval and large interval limits below.  One direct way to see this point is to draw the diagrams for the $W$ functions. The function $W^1_1$ could be defined by moving the integral contour such that
\be \label{contour1}W_1^1=\int_{i y}^{1+i y}du \vartheta_1(u-u_1|i\beta)^{-(1-\frac{k}{n})} \vartheta_1(u-u_2|i\beta)^{-\frac{k}{n}} \vartheta_1(u-(1-\frac{k}{n})u_1-\frac{k}{n}u_2| i\beta)
\ee
The integral contour for $W_1^1$ is along $z=t+iy$ with $t\in[0,1]$, and $y$ can be any number in $[0,\beta]$. The integral does not depend on $y$ such that it can go around singularity. In Fig. \ref{W11} and Fig. \ref{W22} we plot respectively the functions $W_1^1$ and $W_2^2$ with respect to $l=u_2-u_1$, where $r=\frac{k}{n}$.
The $W_1^1$ is finite everywhere (the line $\beta=1/4,r=1/4$ is still finite even though it is out of the picture) but $W_2^2$ is divergent at $l=1$. From these pictures it is clear that the relation $W_2^2=i\beta W_1^1$ suggested in \cite{Datta} is wrong. In the equation (3.14) the term $|\det W|$ contributes the divergence in the large interval limit.

\begin{figure}[tbp]
\centering
\subfloat[The behavior of $W_1^1(l)$ for different values of $\beta$ and $r=k/n$.]{\includegraphics[width=7cm]{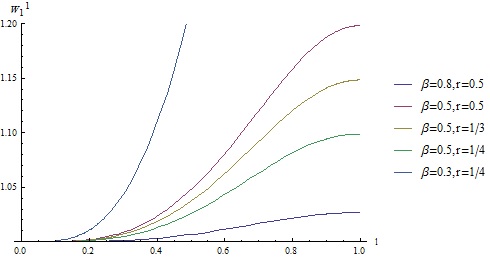}\label{W11}}
\quad
\subfloat[The behavior of $W_2^2(l)$ for different values of $\beta$ and $r=k/n$. ]{\includegraphics[width=7cm]{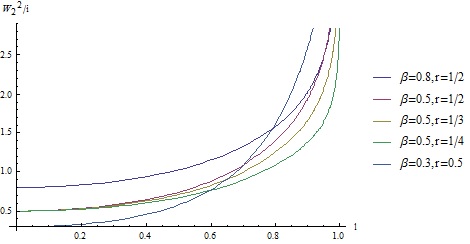}\label{W22}}
\caption{The $W$ functions. It is easy to see that $W_1^1$ is finite, while $W_2^2$ is divergent when in the large interval limit $l \to 1$ }
\end{figure}

\subsection{Small interval expansion}

For short interval, we set $u_1=\frac{1}{2}(1-l)$, and $u_2=\frac{1}{2}(1+l)$. For $l<<1$ we take an expansion with respect to $l$ up to $l^2$. Then we find
\be\label{W11e} W_1^1=\int_0^1due^{\frac{\pi}{\beta}\frac{k}{n}(1-\frac{k}{n})l^2}
\left\{1+\frac{k}{2n}(1-\frac{k}{n})\frac{l^2}{\beta^2}
\left(\frac{\theta_1^{''}(\frac{u-\frac{1}{2}}{i\beta}\mid\frac{i}{\beta})}{\theta_1(\frac{u-\frac{1}{2}}{i\beta}\mid\frac{i}{\beta})}
-\left(\frac{\theta_1^{'}(\frac{u-\frac{1}{2}}{i\beta}\mid\frac{i}{\beta})}{\theta_1(\frac{u-\frac{1}{2}}{i\beta}\mid\frac{i}{\beta})}\right)^2
\right)+O(l^3)\right\}. \ee

Taking second order derivative of the theta function,
we get
\bea\label{expand} \frac{\partial^2}{\partial u^2}\log\vartheta_1(u\mid\tau)&=&
\frac{\vartheta_1^{''}(u\mid\tau)}{\theta_1(u\mid\tau)}-(\frac{\theta_1^{'}(u\mid\tau)}{\theta_1(u\mid\tau)})^2  \notag \\
&=&-\frac{\pi^2}{\sin^2 \pi u}+\sum_{m=1}^{\inf}\frac{4\pi^2e^{2\pi iu}q^m}{(1-e^{2\pi iu}q^m)^2}
+\sum_{m=1}^{\inf}\frac{4\pi^2e^{-2\pi iu}q^m}{(1-e^{-2\pi iu}q^m)^2}.
\eea
Taking (\ref{expand}) into (\ref{W11}), we read
\bea\label{W11s} W_1^1
&=&e^{\frac{\pi}{\beta}\frac{k}{n}(1-\frac{k}{n})l^2}
(1+\frac{1}{2}\frac{k}{n}(1-\frac{k}{n})\frac{l^2}{\beta^2}(-2\pi\beta+O(e^{-\frac{4\pi}{\beta}}))+O(l^3)).
\eea

For $W_2^2$, we can find the result from \cite{Datta}
\be\label{W22s} (W_2^2)^*=i\beta e^{\frac{\pi}{\beta}\frac{k}{n}(1-\frac{k}{n})l^2}[1+O(l^3)+O(e^{-\frac{4\pi}{\beta}})]. \ee
Then (\ref{W11s}) and (\ref{W22s}) are the  small interval expansion for $W_1^1$ and $W_2^2$ at high temperature. Obviously the relation (\ref{relationW}) does not hold in this case.

\subsection{Large interval expansion}

In this subsection, we calculate the high temperature expansion of $W_1^1$ and $W_2^2$ for the large interval. In this case, we may set
\bea &~&u_1=\epsilon_1 \notag \\
&~&u_2=1-\epsilon_2 \eea
with $\epsilon_1,\epsilon_2<<1$.

\begin{figure}[tbp]
\centering
\subfloat[A cycle]{\includegraphics[width=5cm]{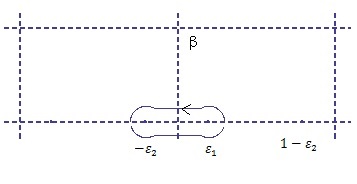}\label{Acycle}}
\quad
\subfloat[B cycle]{\includegraphics[width=5cm]{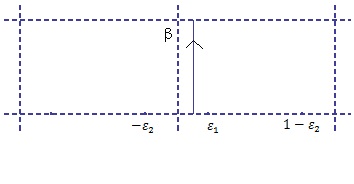}\label{Bcycle}}
\caption{There are two contour integral paths. Under a conformal transformation, they transform into two cycles in the full complex plane with four twist vertex. We have to stress that  the A cycle and B cycle here are different from the ones in Fig. \ref{i1}. }
\end{figure}

For $W_1^1$, we can transform it into a contour integral.
\bea \lefteqn{W_1^1} \notag \\
&=&(\int_0^{u_1}+\int_{u_1}^{u_2}+\int_{u_2}^{1})du\vartheta_1(u-u_1\mid\tau)^{-(1-\frac{k}{n})}\vartheta_1(u-u_2\mid\tau)^{-\frac{k}{n}}
\vartheta_1(u-(1-\frac{k}{n})u_1-\frac{k}{n}u_2\mid\tau) \notag \\
&=&(\int_0^{u_1}+\int_{u_1}^{u_2}+\int_{u_2}^{1})du\vartheta_1(u-u_1\mid\tau)^{-(1-\frac{k}{n})}\vartheta_1(u-u_2\mid\tau)^{-\frac{k}{n}}
\vartheta_1(u-(1-\frac{k}{n})u_1-\frac{k}{n}u_2\mid\tau) \notag \\
&=&\int_{-\epsilon_2}^{\epsilon_1}du\vartheta_1(u-\epsilon_1\mid\tau)^{-(1-\frac{k}{n})}
\vartheta_1(u-(1-\epsilon_2)\mid\tau)^{-\frac{k}{n}}
\vartheta_1(u-(1-\frac{k}{n})\epsilon_1-\frac{k}{n}(1-\epsilon_2)\mid\tau) \notag \\
&=&-\frac{1}{1-e^{-\frac{2\pi ik}{n}}}\oint_A du\vartheta_1(u-\epsilon_1\mid\tau)^{-(1-\frac{k}{n})}
\vartheta_1(u-(1-\epsilon_2)\mid\tau)^{-\frac{k}{n}}
\vartheta_1(u-(1-\frac{k}{n})\epsilon_1-\frac{k}{n}(1-\epsilon_2)\mid\tau). \notag \\
\eea
The contour  is shown in Fig. \ref{Acycle}. The reduction to the second line is because  the integral in $[u_1,u_2]$ vanishes. And by studying the monodromy condition, we can derive the third line.

To obtain a high temperature expansion, we first take an $S$ duality, and expand with respect to $e^{-\frac{2\pi}{\beta}}$,
\bea &~&\vartheta_1(u-\epsilon_1\mid i\beta)^{-(1-\frac{k}{n})}
\vartheta_1(u-(1-\epsilon_2)\mid i\beta)^{-\frac{k}{n}}
\vartheta_1(u-(1-\frac{k}{n})\epsilon_1-\frac{k}{n}(1-\epsilon_2)\mid i\beta) \notag \\
&=&e^{\frac{\pi}{\beta}\frac{k}{n}(1-\frac{k}{n})(1-\epsilon_1-\epsilon_2)^2}
\vartheta_1\left(\frac{u-\epsilon_1}{i\beta}\mid \frac{i}{\beta}\right)^{-(1-\frac{k}{n})}
\vartheta_1\left(\frac{u-(1-\epsilon_2)}{i\beta}\mid \frac{i}{\beta}\right)^{-\frac{k}{n}}
\vartheta_1\left(\frac{u-(1-\frac{k}{n})\epsilon_1-\frac{k}{n}(1-\epsilon_2)}{i\beta}\mid \frac{i}{\beta}\right) \notag \\
&=&(-1)e^{\frac{\pi}{\beta}\frac{k}{n}(1-\frac{k}{n})(1-\epsilon_2)^2}e^{-\frac{k}{n}\pi i}e^{\frac{2\pi u}{\beta}\frac{k}{n}}
(e^{\frac{2\pi u}{\beta}}-1)^{-(1-\frac{k}{n})}(e^{\frac{2\pi u}{\beta}}-e^{-\frac{2\pi\epsilon_2}{\beta}})^{-\frac{k}{n}}
\left\{1-e^{\frac{2\pi u}{\beta}}e^{-\frac{2\pi}{\beta}\frac{k}{n}}e^{\frac{2\pi}{\beta}\frac{k}{n}\epsilon_2} \right. \notag \\
&~&\left.-e^{-\frac{2\pi u}{\beta}}e^{-\frac{2\pi}{\beta}(1-\frac{k}{n})}e^{-\frac{2\pi}{\beta}\frac{k}{n}\epsilon_2}
+O(e^{-\frac{2\pi}{\beta}})\right\}.
  \eea
In the second equation we use the translation relation
\bea \vartheta_1(u+1\mid\tau)&=&-\vartheta_1(u\mid\tau), \notag \\
\vartheta_1(u+\tau\mid\tau)&=&-\frac{1}{e^{2\pi iu}e^{\pi i\tau}}\vartheta(u\mid \tau),\eea
and in the last equation, we set $\epsilon_1=0$ for simplicity. Actually, in the final result it is always the combination $\epsilon_1+\epsilon_2$ appearing in the discussion, so we lose nothing by just keeping $\epsilon_2$.

Under a conformal transformation
\be z=e^{\frac{2\pi u}{\beta}}, \ee
we find that
\bea
W_1^1&=&\frac{1}{1-e^{-2\pi i\frac{k}{n}}}\oint_A dz\frac{\beta}{2\pi}
e^{\frac{\pi}{\beta}\frac{k}{n}(1-\frac{k}{n})(1-\epsilon_2)^2}e^{-\frac{k}{n}\pi i}z^{-(1-\frac{k}{n})}
(z-1)^{-(1-\frac{k}{n})}(z-e^{-\frac{2\pi\epsilon_2}{\beta}})^{-\frac{k}{n}}\cdot\notag \\
&~&\left\{1-ze^{-\frac{2\pi}{\beta}\frac{k}{n}}e^{\frac{2\pi}{\beta}\frac{k}{n}\epsilon_2}
-z^{-1}e^{-\frac{2\pi}{\beta}(1-\frac{k}{n})}e^{-\frac{2\pi}{\beta}\frac{k}{n}\epsilon_2}+O(e^{-\frac{2\pi}{\beta}})\right\}.
\eea

The integral over $z$ goes around $1$ and $x\equiv e^{-\frac{2\pi\epsilon_2}{\beta}}$, and leads to hyper-geometric functions  \cite{Dixon0}\cite{Cardy1}. By the integral form of hyper-geometric function
\be F(\alpha,\beta,\gamma,z)=\frac{\Gamma(\gamma)}{\Gamma(\beta)\Gamma(\gamma-\beta)}
\int_0^1t^{\beta-1}(1-t)^{\gamma-\beta-1}(1-zt)^{-\alpha}dt, \ee
we have
\bea &~&\oint_A dzz^{-(1-\frac{k}{n})}(z-x)^{-\frac{k}{n}}(z-1)^{-(1-\frac{k}{n})}=2\pi iF(1-\frac{k}{n},\frac{k}{n},1,1-x), \notag \\
&~&\oint_A dzz^{\frac{k}{n}}(z-x)^{-\frac{k}{n}}(z-1)^{-(1-\frac{k}{n})}=2\pi iF(-\frac{k}{n},\frac{k}{n},1,1-x), \notag \\
&~&\oint_A dzz^{-(2-\frac{k}{n})}(z-x)^{-\frac{k}{n}}(z-1)^{-(1-\frac{k}{n})}=2\pi iF(2-\frac{k}{n},\frac{k}{n},1,1-x).  \eea
With all these relations, we have
\bea\label{W11l} W_1^1&=&\frac{\beta}{2\sin\frac{\pi k}{n}}e^{\frac{\pi}{\beta}\frac{k}{n}(1-\frac{k}{n})(1-\epsilon_2)^2}
\left\{F(1-\frac{k}{n},\frac{k}{n},1,1-x)-e^{-\frac{2\pi}{\beta}\frac{k}{n}}x^{-\frac{k}{n}}F(-\frac{k}{n},\frac{k}{n},1,1-x) \right.\notag \\
&~&\left.-e^{-\frac{2\pi}{\beta}(1-\frac{k}{n})}x^{\frac{k}{n}}F(2-\frac{k}{n},\frac{k}{n},1,1-x)+O(e^{-\frac{2\pi}{\beta}})\right\}. \eea

We can compute $W_2^2$ in the same way. First, we notice that
\bea\label{W22lq}
(W_2^2)^*&=&\int_0^{i\beta}du\vartheta_1(u-u_1\mid i\beta)^{-\frac{k}{n}}
\vartheta_1(u-u_2\mid i\beta)^{-(1-\frac{k}{n})}
\vartheta_1(u-\frac{k}{n}u_1-(1-\frac{k}{n})u_2) \notag \\
&=&\oint_B\frac{\beta}{2\pi}dze^{\frac{\pi}{\beta}\frac{k}{n}(1-\frac{k}{n})(1-\epsilon_2)^2}
e^{\pi i\frac{k}{n}}z^{-\frac{k}{n}}(z-1)^{-\frac{k}{n}}(z-x)^{-(1-\frac{k}{n})}
\left\{1-zx^{-(1-\frac{k}{n})}e^{-\frac{2\pi}{\beta}(1-\frac{k}{n})}\right. \notag \\
&~&\left.-z^{-1}x^{(1-\frac{k}{n})}e^{-\frac{2\pi}{\beta}\frac{k}{n}}
+O(e^{-\frac{2\pi}{\beta}})\right\}.
\eea
Similarly we set $\epsilon_1=0$ for simplicity. In the last equality, we take a coordinate transformation into the full complex plane. The integral
 contour goes around $0$ and $x$. As before, the integrals can be expressed in terms of  hypergeometric functions. The first two terms give
\bea &~&\oint dzz^{-\frac{k}{n}}(z-1)^{-\frac{k}{n}}(z-x)^{-(1-\frac{k}{n})}=2\pi ie^{-\pi i\frac{k}{n}}
F(\frac{k}{n},1-\frac{k}{n},1,x), \notag \\
&~& \oint dzz^{1-\frac{k}{n}}(z-1)^{-\frac{k}{n}}(z-x)^{-(1-\frac{k}{n})}=2\pi i(1-\frac{k}{n})e^{-\pi i\frac{k}{n}}
xF(\frac{k}{n},2-\frac{k}{n},2,x). \eea
For the last term, we need to take a transformation $\tilde{z}=\frac{1}{z}$
\bea &~&\oint dzz^{-1-\frac{k}{n}}(z-1)^{-\frac{k}{n}}(z-x)^{-(1-\frac{k}{n})} \notag \\
&=&\oint\frac{d\tilde{z}}{\tilde{z}^2}\tilde{z}^{1+\frac{k}{n}}(\frac{1}{\tilde{z}}-1)^{-\frac{k}{n}}(\frac{1}{\tilde{z}}-x)^{-(1-\frac{k}{n})} \notag \\
&=&2\pi i\frac{k}{n}e^{-\pi i\frac{k}{n}}F(1-\frac{k}{n},1+\frac{k}{n},2,x) \eea
Taking these relations into (\ref{W22lq}), we finally obtain
\bea\label{W22l} (W_2^2)^*&=&\beta i\exp\left(\frac{k}{n}(1-\frac{k}{n})\frac{\pi}{\beta}(1-\epsilon_2)^2\right)
\left\{F(\frac{k}{n},1-\frac{k}{n},1,x)-(1-\frac{k}{n})e^{-\frac{2\pi}{\beta}(1-\frac{k}{n})}x^{\frac{k}{n}}F(\frac{k}{n},2-\frac{k}{n},2,x) \right. \notag \\
&~&\left.-\frac{k}{n}e^{-\frac{2\pi}{\beta}\frac{k}{n}}x^{1-\frac{k}{n}}F(1-\frac{k}{n},1+\frac{k}{n},2,x)
+O(e^{-\frac{2\pi}{\beta}})\right\}. \eea

In short,  the large interval expansion of $W_1^1$ and $W_2^2$ at high temperature  are given by (\ref{W11l}) and (\ref{W22l}) respectively. The relation (\ref{relationW}) does not hold in this case neither.

\subsection{Small and large interval limit for the W function}

In this section, we just list some results from \cite{Chen:2015cna} for the smalll and large interval limits. For the small interval, when $u_1=u_2$
\be\label{sW} W_1^{1(k)}=1~~~\bar{W}_2^{2(k)}=i\beta. \ee
For the large interval
\be\label{lW1} W_1^{1(k)}=-\frac{1}{2\sin\pi\frac{k}{n}}\eta(\tau)^{-3}\vartheta_1(-\frac{k}{n}\mid\tau)+O(u_1-(u_2-1)), \ee
and
\be\label{lW2} \bar{W}_2^{2(k)}= i\frac{\sin\frac{\pi k}{n}}{\pi}\eta(\tau)^{-3}\vartheta_1(-(1-\frac{k}{n})|\tau)
\log~(u_1-(u_2-1))+O((1-u_2+u_1)^0). \ee

\begin{figure}[tbp]
\centering
\subfloat[The  function  $W_1^1(l=1, \beta)$ for different $r$. This is from the contour integral (\ref{contour1}).]{\includegraphics[width=7cm]{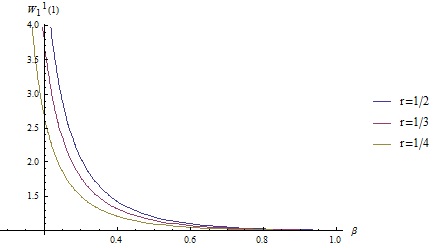}\label{W11(1)1}}
\quad
\subfloat[The function $W_1^1(l=1, \beta)$ from analytic expression (\ref{lW1}). ]{\includegraphics[width=7cm]{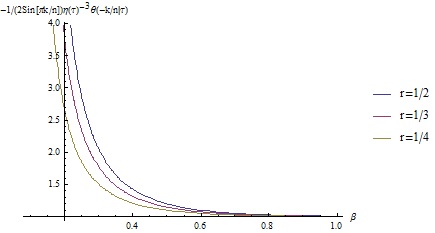}\label{W11(1)2}}
\caption{The comparison of $W_1^1(l=1,\beta)$ between the direct contour integral (\ref{contour1}) and analytic expression (\ref{lW1}).  }
\end{figure}

\begin{figure}
  \centering
  \includegraphics[width=7cm]{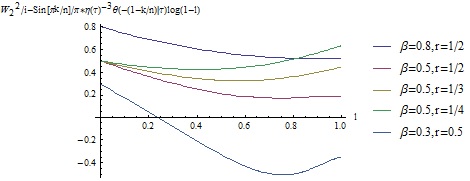}\\
  \caption{Subtract the logarithmic divergence from $(W_2^2)^*$. After subtraction, the behavior of the function is regular and finite. }\label{W22(1)}
\end{figure}

In Fig. \ref{W11(1)1} and  Fig. \ref{W11(1)2}, we plot the dependence of $W_1^1(l=1)$ on $\beta$ for different values of $r$.  Both diagrams are in good agreement.  This supports our calculation for $W_1^1$. Our calculation corrects the coefficient in (B.51) in \cite{Datta}.
The function $(W_2^2)^*$ is divergent at $l=1$. From (A.26) we have
\be (W_2^2)*\sim i \frac{sin\frac{\pi k}{n}}{\pi}\eta(\tau)^{-3}\vartheta_1(-(1-\frac{k}{n}))|\tau)
\log(1-l). \ee
In Fig. \ref{W22(1)} we subtract the divergent term proportional to $\log(1-l)$ and plot the expression $\frac{1}{i}(W_2^2)^*-\frac{sin\frac{\pi k}{n}}{\pi} \eta(\tau)^{-3}
\vartheta_1(-(1-\frac{k}{n}|\tau) \log(1-l) $ for different $\beta$ and $r$. We find that the subtracted function is always finite at $l=1$. Therefore, we conclude that our treatment on the W functions is consistent with the numerical computation.

\subsection{Recursion relations for hypergeometric function}

Here we list some recursion relations, which are useful in our calculation
\begin{align}\label{recursion} \frac{(\gamma-\alpha)_n(\gamma-\beta)_n}{(\gamma)_n}(1-z)^{\alpha+\beta-\gamma-n}
F(\alpha,\beta,\gamma+n,z)&=\frac{d^n}{dz^n}[(z-z)^{\alpha+\beta-\gamma}F(\alpha,\beta,\gamma,z)], \notag \\
 (\gamma-\beta-1)F+\beta F(\beta+1)-(\gamma-1)F(\gamma-1)&=0, \notag \\
\gamma(\alpha-(\gamma-\beta)z)F-\alpha\gamma(1-z)F(\alpha+1)+(\gamma-\alpha)(\gamma-\beta)zF(\gamma+1)&=0, \notag \\
 (\gamma-1)F(\gamma-1)-\alpha F(\alpha+1)-(\gamma-\alpha-1)F&=0, \notag \\
\gamma(1-z)F-\gamma F(\beta-1)+(\gamma-\alpha)zF(\gamma)&=0. \end{align}
\end{appendix}



\end{document}